\documentclass[runningheads,a4paper]{llncs}

        \makeatletter
        \renewcommand*\l@author[2]{}
        \renewcommand*\l@title[2]{}
        \makeatletter

\usepackage[T1]{fontenc}
\usepackage{stmaryrd}
\usepackage{algorithm}
\usepackage{algorithmic}
\usepackage{graphicx}
\usepackage{floatflt}
\usepackage{wrapfig}
\usepackage{pstricks}
\usepackage{rotating}
\usepackage{subfigure}
\usepackage{amsmath}
\usepackage{amssymb}
\usepackage{theorem}
\usepackage[active]{srcltx} 
\usepackage{amsmath,amssymb}
\usepackage{booktabs}
\usepackage{xcolor}
\usepackage{hyperref}
\usepackage{listings}
\usepackage{enumerate}
\usepackage[justification=RaggedRight, singlelinecheck=false]{caption}
\usepackage{marvosym}

\newcommand{\QED}{\hspace*{\fill}$\Box$}

\newcommand{\T}{{\cal T}}
\newcommand{\K}{{\cal K}}
\newcommand{\A}{{\cal A}}
\newcommand{\B}{{\cal B}}
\newcommand{\swedge}{\sqcap}
\newcommand{\bigswedge}{{\bigsqcap}}

\newcommand{\ignore}[1]{}

\newenvironment{enumerate-} 
{\begin{enumerate}
    
   \setlength{\parskip}{-1ex}              
   \setlength{\itemsep}{1.5ex}             
}
{
 \end{enumerate}
}

\newtheorem{thm}{Theorem}
\newtheorem{lem}[thm]{Lemma}

\newtheorem{cor}[thm]{Corollary}
\newtheorem{prop}[thm]{Proposition}

\begin{document}

\pagestyle{headings}
\title{On $P$-Interpolation in Local Theory Extensions and
Applications to the Study of  Interpolation in the
  Description Logics ${\cal EL}, {\cal EL}^+$}

\titlerunning{$P$-interpolation in local theory extensions and
  applications to ${\cal EL}, {\cal EL}^+$}
\author{Dennis Peuter \and Viorica
  Sofronie-Stokkermans \and Sebastian Thunert} 
\authorrunning{Dennis Peuter, Viorica Sofronie-Stokkermans, Sebastian Thunert}
\institute{University of Koblenz, Koblenz, Germany\\
$\{$dpeuter,sofronie$\}$@uni-koblenz.de,  s.thunert@gmx.de}

\maketitle

\begin{abstract}
We study the problem of $P$-interpolation, where $P$ is a set of
binary predicate symbols, for certain classes of 
local extensions of a base theory. 
For computing the $P$-in\-ter\-po\-la\-ting terms, we use a hierarchic
approach: This allows us to compute the interpolating terms using a
method for computing interpolating terms in the base theory. 
We use these results for 
proving $\leq$-interpolation in classes of semilattices with monotone
operators; 
we show, by giving a counterexample, that $\leq$-interpolation
  does not hold 
if by ``shared'' symbols we mean just the {\em common} symbols.
We use these results for the study of $\sqsubseteq$-interpolation 
in the description logics ${\cal EL}$ and ${\cal EL}^+$.
\end{abstract}

\section{Introduction}
In this paper we study the problem of $P$-interpolation, 
a problem strongly related to interpolation w.r.t.\ logical
theories. The problem can be formulated as follows: 

\smallskip
\noindent 
Let ${\mathcal T}$ be a theory, $A$ and $B$ be 
conjunctions of ground literals in the signature of ${\mathcal T}$, possibly 
with additional constants, $P$  a set of binary predicate symbols in
the signature of $\T$, 
$R \in P$, $a$ a constant occurring
in $A$ and $b$ a constant occurring in $B$.
Assume that $A \wedge B \models_{\cal T} a R b$. Can we find a ground 
term $t$ containing only constants and function symbols ``shared'' by 
$A$ and $B$, such that $A \wedge B \models_{\cal T} a R t ~\wedge~ t R
b$?

\smallskip
\noindent Interpolation has been studied in classical and 
non-classical logics and in extensions and combinations of theories; 
and is very important in program verification and also in the area of
description logics. 
The first algorithms for interpolant generation in program 
verification required explicit constructions and ``separations'' 
of proofs \cite{Krajicek97,McMillanProver04}. 
In \cite{kovacs-voronkov} interpolants are
computed using variants of resolution.
For certain theories,
the ``separation'' of proofs relied on the
possibility of ``separating'' atoms, i.e.\ 
on $P$-interpolation, where $P$ is a set of binary predicate symbols. 
Equality interpolation is used in
\cite{Yorsh-Musuvathi} for devising an interpolation method in
combinations of theories with disjoint signatures.  
In  \cite{Sofronie-ijcar-06,Sofronie-lmcs} and \cite{rybal-sofronie}, for instance, we consider
interpolation problems in certain
classes of extensions $\T_0 \cup \K$ of a base theory $\T_0$ and 
use a hierarchical approach to compute
interpolants. The method relies on the $P$-interpolation property of the base
theory $\T_0$.
In most of the applications we considered, 
$P$ contains the equality predicate $\approx$ or a predicate $\leq$ with the
property that in all models of $\T_0$, the interpretation of $\leq$ is
a partial ordering. 

\smallskip 
\noindent Since at that time our main interest was the study of {\em interpolation
problems}, in  \cite{Sofronie-ijcar-06,Sofronie-lmcs} and 
\cite{rybal-sofronie} $P$-interpolation is only used in order to help in giving methods
for interpolation and not as a goal in itself. 
However, in several papers in the area of description logics (cf. e.g.\
\cite{ten-Cate-et-al-13,kr-22}) when defining the notion of 
interpolation in description logics the authors define in fact a
notion of $\sqsubseteq$-interpolation.  In \cite{kr-22} (Theorem~4) 
it is proved that ${\cal EL}^+$ allows interpolation (in fact, the notion of 
$\sqsubseteq$-interpolation mentioned above) for {\em safe} role
inclusions -- this is related to the notion of ``sharing''
considered in \cite{Sofronie-lmcs}, cf.\ also Section~\ref{p-int}. 
The proof technique in \cite{kr-22} uses simulations. 
In this paper, we analyze the property of $P$-interpolation in theory 
extensions, propose a method for solving it based on
hierarchical reasoning 
and formulate the $\sqsubseteq$-interpolation problem for ${\cal EL}$ 
and ${\cal EL}^+$ as a $\leq$-interpolation problem in a theory of semilattices
with operators.

\smallskip
\noindent We first studied $\leq$-interpolation in \cite{peuter-sofronie}
in the context of description logics; 
the $\sqsubseteq$-interpolating concept descriptions were 
regarded as a form of ``high-level'' explanations.
In this paper we further extend the work  in \cite{peuter-sofronie}.
The general approach we propose opens the possibility of applying 
similar methods to more general classes of non-classical logics
(including e.g.\ substructural logics or the 
logics with monotone operators studied in  \cite{Ihlemann-Sofronie-ismvl,sofronie-ihlemann-ismvl-07}) or in 
verification (to consider more general 
theory extensions than those with uninterpreted function symbols analyzed 
in \cite{rybal-sofronie}). 
The main results can be summarized as follows:  
\begin{itemize}
\item We propose variants of the definitions of convexity,
  $P$-interpolation and Beth definability relative to a subsignature.
\item We describe a hierarchical $P$-interpolation method in 
certain classes of local theory extensions.  
\item We illustrate the applicability of these results to 
prove that certain classes of semilattices with monotone operators 
have the property of $\leq$-interpolation for a certain interpretation
of ``shared'' function symbols. 
\item We show, by giving a counterexample, that $\leq$-interpolation
  does not hold if by ``shared'' symbols we mean just the {\em common} symbols.
\item We indicate how these results can be used to prove or disprove
  various notions of interpolation for the  description logics ${\cal
    EL}$ and ${\cal EL}^+$. 
\end{itemize}
{\em Structure of the paper:}\/ 
In Section~\ref{prelim} basic notions in logic are briefly introduced, and some results
on convex theory and the link between two versions of 
$\approx$-interpolation and corresponding versions of Beth
definability -- needed later in the paper -- are proved. In Section~\ref{local} we introduce some
results on local theory extensions needed in the paper.   
In Section~\ref{p-int} we identify classes of local theory extensions
allowing $P$-interpolation and propose a hierarchical method of computing
$P$-interpolants. This is used in
Section~\ref{semilattices} to study the existence of 
$\leq$-interpolation in classes of semilattices with monotone operators.
In Section~\ref{el} we use the links between the theory of
semilattices with operators and the description logics ${\cal EL}$ and
${\cal EL}^+$, and show how the results can be used in
the study of these logics. Section~\ref{conclusions} contains the 
conclusions and some plans for future work.

\smallskip
\noindent 
This paper is the extended version of \cite{peuter-sofronie-thunert-cade-23}: it provides
details of proofs and additional examples. 



\vspace{-2mm}
\section*{Table of Contents}

\contentsline {section}{\numberline {1}Introduction}{1}{section.1.1}
\contentsline {section}{\numberline {2}Preliminaries}{3}{section.1.2}
\contentsline {subsection}{\numberline {2.1}Convexity and $P$-convexity}{4}{subsection.1.2.1}
\contentsline {subsection}{\numberline {2.2}Equality interpolation, $R$-interpolation}{4}{subsection.1.2.2}
\contentsline {subsection}{\numberline {2.3}Beth definability}{6}{subsection.1.2.3}
\contentsline {section}{\numberline {3}Local Theory Extensions}{7}{section.1.3}
\contentsline {subsection}{\numberline {3.1}Hierarchical reasoning in local theory extensions}{8}{subsection.1.3.1}
\contentsline {subsection}{\numberline {3.2}Partial structures, weak validity}{9}{subsection.1.3.2}
\contentsline {subsection}{\numberline {3.3}Flat and linear clauses; Flattening and purification}{9}{subsection.1.3.3}
\contentsline {subsection}{\numberline {3.4}Recognizing locality and $\Psi $-locality}{10}{subsection.1.3.4}
\contentsline {section}{\numberline {4}$R$-interpolation in local theory extensions}{12}{section.1.4}
\contentsline {section}{\numberline {5}Example: Semilattices with monotone operators}{17}{section.1.5}
\contentsline {subsection}{\numberline {5.1}The theory ${\sf SLat}$ of semilattices}{17}{subsection.1.5.1}
\contentsline {subsection}{\numberline {5.2}Semilattices with operators}{20}{subsection.1.5.2}
\contentsline {section}{\numberline {6}Applications to ${\cal EL}$ and ${\cal EL}^+$-Subsumption}{24}{section.1.6}
\contentsline {subsection}{\numberline {6.1}Algebraic semantics for ${\cal EL}, {\cal EL}^+$ and $\sqsubseteq $-interpolation}{25}{subsection.1.6.1}
\contentsline {subsection}{\numberline {6.2}Example: $\sqsubseteq $-Interpolation for ${\cal EL}^+$}{26}{subsection.1.6.2}
\contentsline {subsection}{\numberline {6.3}Prototype implementation}{30}{subsection.1.6.3}
\contentsline {section}{\numberline {7}Conclusions and future work}{30}{section.1.7}

\vspace{-2mm}
\section{Preliminaries}

\label{prelim}

We assume known standard definitions from first-order logic  
such as $\Pi$-structures, models, homomorphisms, logical entailment,  
satisfiability, unsatisfiability.  

\smallskip
\noindent 
We consider signatures of the form $\Pi = (\Sigma, {\sf Pred})$, 
where $\Sigma$ is a family of function symbols and ${\sf Pred}$
a family of predicate symbols. If $C$ is a fixed countable set of fresh constants, we denote by 
$\Pi^C$ the extension of $\Pi$ with constants in $C$.  
We denote ``falsum'' with $\perp$. 

\smallskip
\noindent A theory $\T$ is described by a set of closed formulae 
(the axioms of the theory). We call a theory axiomatized by a set of
universally quantified equations an {\em equational theory}. 
We denote by 
${\sf Mod}({\cal T})$ the set of all models of ${\cal T}$. 

\smallskip
\noindent If $F$ and $G$ are formulae we 
write $F \models G$ to express the fact that every model of $F$ 
is a model of $G$; if 
$\T$ is a theory, we write 
$F \models_{\cal T} G$ to express the fact that every model of $F$ 
which is also a model of 
$\T$ is a model of $G$. 
The definitions can be extended in a natural way to the case when 
$F$ is a set of formulae; in this case, $F \models_{\cal T} G$ if and only if 
${\cal T} \cup F \models G$. 
$F \models \perp$ means that $F$ is
unsatisfiable; $F \models_{\T} \perp$ means that there is no model of
$\T$ which is also a model of $F$. If there is a model of $\T$ which is also a
model of $F$ we say
that $F$ is $\T$-consistent.

\smallskip
\noindent In the next sections we introduce notions of convexity of a theory
with respect to a subset of the predicate symbols, and establish links 
between two versions of $\approx$-interpolation and corresponding 
versions of Beth definability, results which are needed later in the paper

\vspace{-2mm}
\subsection{Convexity and $P$-convexity} 
We can define a notion of convexity w.r.t.\ a subset $P$ of the set of predicates. 
\begin{definition}
A theory ${\cal T}$ with signature $\Pi = (\Sigma, {\sf Pred})$ is 
{\em convex}\/ with respect to a subset $P$ of ${\sf Pred}$  (which may include also equality
$\approx$) if for all conjunctions $\Gamma$ of ground 
$\Pi^C$-atoms (with additional constants in a set $C$), 
relations $R_1, \dots, R_m \in P$ and tuples of $\Pi^C$-terms of 
corresponding arity 
${\overline t}_1, \dots, {\overline t}_m$ such that $\Gamma
  \models_{\T} \bigvee_{i = 1}^m R_i({\overline t}_i)$ 
there exists $i_0 \in  \{ 1, \dots, m \}$ such that 
$\Gamma \models_{\T} R_{i_0}({\overline t}_{i_0})$.
\end{definition}
We will call a theory $\T$ {\em convex} if it is ${\sf Pred} \cup \{\approx\}$-convex. 
The following result is well-known (cf.\ e.g.\
\cite{BurrisSankappanavar,Hodges,Tinelli03}): 
\begin{thm}
Let $\T$ be a theory and let ${\sf
    Mod}({\cal T})$ be the class of models of $\T$. 
\begin{enumerate}[(i)]
\item If ${\sf
    Mod}({\cal T})$ is closed under direct products then $\T$ is
  convex.
\item If $\T$ is a universal theory and $\T$ is convex, 
  then $\T$ has an axiomatization given by Horn clauses, hence ${\sf
    Mod}({\cal T})$ is closed under direct products. \!\!
\end{enumerate}
\label{thm-convex}
\end{thm}
\begin{cor} Let $\T_1$, $\T_2$ be two theories with signatures
  $\Pi_1, \Pi_2$. 
If ${\sf Mod}(\T_1)$ and ${\sf Mod}(\T_2)$ are closed under
direct products, then $\T_1 \cup \T_2$ is convex. 
\label{cor-convex}
\end{cor}
{\em Proof.} ${\sf Mod}(\T_1 \cup \T_2) = \{ {\cal A}  \mid
{\cal A} \text{ is a } (\Pi_1 \cup
\Pi_2)\text{-structure with } {{\cal A}}_{|\Pi_1} \in {\sf
  Mod}(\T_1) \text{ and }$ ${{\cal A}}_{|\Pi_2} \in {\sf
  Mod}(\T_2) \}$.  Let ${\cal A}_i \in {\sf Mod}(\T_1 \cup
\T_2), i \in I$. This is the case iff ${{\cal A}_i}_{| \Pi_1} \in
{\sf Mod}(\T_1)$ and ${{\cal A}_i}_{| \Pi_2} \in
{\sf Mod}(\T_2)$ for all $i \in I$. Taking into account that, due to
the definition of a direct product of structures, 
$\big( \prod_{i \in I} {\cal  A}_i \big)_{|\Pi_j} =
 \prod_{i \in I} \big( {{\cal A}_i}_{|\Pi_j} \big)$ for $j = 1, 2$, and the fact that ${\sf
Mod}(\T_1)$ 
and ${\sf Mod}(\T_2)$ are closed under direct products, 
it follows that $\prod_{i \in I} {\cal A}_i \in {\sf Mod}(\T_1 \cup
\T_2)$, so ${\sf Mod}(\T_1 \cup \T_2)$ is closed under products. 
From Theorem~\ref{thm-convex} it follows that $\T_1 \cup \T_2$ is convex. \QED

\begin{cor}
Let $\T_1$ and $\T_2$ be convex universal theories. Then $\T_1
\cup \T_2$ is convex.  
\end{cor}
{\em Proof:} 
By Theorem~\ref{thm-convex}, since $\T_i$ is a universal theory and
convex,  ${\sf
    Mod}({\cal T}_i)$ is closed under direct products, for $i = 1, 2$.
By Corollary~\ref{cor-convex} it follows that ${\sf Mod}(\T_1 \cup
\T_2)$ is closed under products, hence $\T_1 \cup \T_2$ is also
convex. \QED

\medskip
\noindent 
In particular, every extension of a convex universal theory $\T_0$ with a
set of new function symbols axiomatized by a set $\K$ of Horn clauses
is convex. 

\vspace{-2mm}
\subsection{Equality interpolation, $R$-interpolation}
We now define various versions of interpolation w.r.t.\ certain
predicates. 

\begin{definition}
We say that a convex theory $\T$ has the equality interpolation property if 
for every conjunction of ground $\Pi^C$-literals 
$A(\overline{c}, \overline{a_1}, a)$ and 
$B(\overline{c}, \overline{b_1}, b)$,  
if $A \wedge B \models_{\cal T} a \approx b$
then there exists a term $t(\overline{c})$ containing only the 
constants $\overline{c}$ shared by $A(\overline{c}, \overline{a_1}, a)$ and 
$B(\overline{c}, \overline{b_1}, b)$ such that $A \wedge B \models_{\cal T} a \approx
t(\overline{c}) ~\wedge~ t(\overline{c}) \approx b$. 
\end{definition}
\noindent Sometimes, the theories and theory extensions we study
contain interpreted symbols in a set $\Pi_0 = (\Sigma_0, {\sf Pred})$ and non-interpreted
function symbols in a set $\Sigma_1$. The classical definition for
equality interpolation for a theory $\T$ mentioned above allows the
term $t(\overline{c})$ to contain
all function symbols in the signature of $\T$ -- these symbols are in
this case all seen as being interpreted. If we distinguish between
interpreted and uninterpreted functions we might allow 
the intermediate term $t(\overline{c})$ to contain any interpreted
function symbols, but   
require that $t(\overline{c})$ contains only ``shared''
uninterpreted functions 
and common constants. 

\smallskip
\noindent 
If $\Sigma_A$ and $\Sigma_B$ 
are the uninterpreted function symbols occurring in $A$ resp. $B$, and
$\Theta$ is a closure operator,  by ``shared'' uninterpreted functions we can 
mean: 
\begin{itemize}
\item {\em Intersection-shared symbols}:  ${\bigcap}$-${\sf Shared}(A, B) =
  \Sigma_A \cap \Sigma_B$, or  
\item {\em $\Theta$-shared symbols}: 
$\Theta$-${\sf Shared}(A, B) = \Theta(\Sigma_A) \cap \Theta(\Sigma_B)$.
\end{itemize}
\begin{example}
Let $\T = \T_0 \cup \K$ be the extension of a theory $\T_0$ with set of
interpreted function symbols $\Sigma_0$ with a set $\K$
of clauses containing new uninterpreted function symbols in a set
$\Sigma_1$. 
\begin{itemize}
\item If $A$
and $B$ are sets of atoms in the signature of $\T$ containing additional
constants in a set $C$ and uninterpreted function symbols
$\Sigma_A$, $\Sigma_B$ then the {\em intersection-shared}
uninterpreted function symbols of $A$ and $B$ are $\Sigma_A \cap \Sigma_B$. 
\item For every $f, g \in \Sigma_1$ we define $f \sim_{\K} g$ iff there
exists $C \in \K$ such that $f, g$ both occur in $C$. Let $\sim^*_{\K}$ be
the equivalence relation on $\Sigma_1$ induced by $\sim_{\K}$.

Let $\Theta_{\K}$ be defined for
    every $\Sigma \subseteq \Sigma_1$ by $\Theta_{\K}(\Sigma) =
    \bigcup_{f \in \Sigma} \{ g \in \Sigma_1 \mid f \sim^*_{\K} g \}.$ 
Then the $\Theta_{\K}$-shared symbols are  $\Theta_{\K}(\Sigma_A) \cap \Theta_{\K}(\Sigma_B)$.

In particular, if $A$ contains a
function symbol $f$ and $B$ contains a symbol  $g$ such that $f,
g$ occur both in a clause in $\K$, then $f$ and $g$ are considered to
be $\Theta_\K$-shared by $A$, $B$. \QED
\end{itemize}
\label{ex-shared}
\end{example}
We are interested in similar properties for other
binary relations than $\approx$.
We define an $R$-interpolation property, 
where $R$ is a binary predicate symbol in $\Pi$.
\begin{definition} 
Let $R \in {\sf Pred} \cup \{ \approx \}$ be a binary predicate symbol. 
An $\{R\}$-convex theory $\T$ with uninterpreted symbols $\Sigma_1$ 
has the {\em $R$-interpolation property} 
if  for all conjunctions of ground atoms 
$A(\overline{c}, \overline{a_1}, a)$ and 
$B(\overline{c}, \overline{b_1}, b)$, 
if $A \wedge B \models_{\cal T} a R b$
then there exists a term $t(\overline{c})$ 
containing only common constants $\overline{c}$ and only ``shared''
uninterpreted symbols in $\Sigma_1$ such that $A \wedge B \models_{\cal T} a R
t(\overline{c}) ~\wedge~ t(\overline{c}) R b$. 
%

If $P \subseteq {\sf Pred}$, we say that a theory has the {\em
  $P$-interpolation property} 
if it has the $R$-interpolation property 
for every $R \in P$.  
\label{r-int}
\end{definition} 
In Section~\ref{semilattices} we give examples of theories with this property
and show that a theory may not have the
$R$-interpolation property for a predicate symbol $R$ 
if we use the notion of {\em intersection-shared symbols}, but 
has the $R$-interpolation property if we consider the less restrictive
notion of {\em $\Theta$-shared symbols} for a suitably defined closure
operator $\Theta$.

\vspace{-2mm}
\subsection{Beth definability} 
Let $\T$ be a theory with signature
$\Pi = (\Sigma_0 \cup \Sigma_1, {\sf Pred})$, where the function
symbols in $\Sigma_0$ are regarded as interpreted function symbols and the
function symbols in $\Sigma_1$ are regarded as uninterpreted function
symbols, and let $C$ be a set of additional constants. We define 
a notion of Beth definability relative to a subset $\Sigma_S \subseteq \Sigma_1 \cup C$
of non-interpreted function symbols and constants (similar to the one
introduced in \cite{ten-Cate-et-al-13}), which
we refer to as $\Sigma_S$-Beth definability. 
 
\noindent Let $\Sigma_S \subseteq \Sigma_1 \cup C$, let $\Sigma_r
= \Sigma_1 \backslash \Sigma_S$, and let 
$\Pi' = (\Sigma_0 \cup (\Sigma_S \cap \Sigma_1) \cup \Sigma'_r, {\sf
  Pred})$, where $\Sigma'_r = \{ f' \mid f \in \Sigma_1
\backslash \Sigma_S \}$ 
is the signature obtained by replacing all uninterpreted function symbols in $\Sigma_1$ which are not
in $\Sigma_S$ with new primed copies. 
If $\phi$ is a $\Pi^C$-formula, we will denote by $\phi'$ the formula obtained from
$\phi$ by replacing all uninterpreted function symbols in $\Sigma_1
\backslash \Sigma_S$ and all constants in $C \backslash \Sigma_S$
with distinct, primed versions. The interpreted function symbols and
the uninterpreted function symbols and constants in $\Sigma_S$ are not
changed. 
We regard the theory $\T$ as a set of formulae; 
let $\T' := \{ \phi' \mid \phi \in \T \}$.
 
\begin{definition}
Let $A$ be a conjunction of ground $\Pi^C$-literals, and $a \in C$.
\begin{itemize}
\item We say that  $a$ is {\em implicitly  defined} by $A$
w.r.t. $\Sigma_S$ and $\T$ if,  with the notations introduced before,  
$A \wedge A' \models_{\T \cup \T'} a \approx a'.$
\item We say that $a$ is {\em explicitly defined} by $A$ w.r.t.\ $\Sigma_S$
and $\T$ 
if there exists a term $t$ containing only symbols in $\Sigma_0, {\sf Pred}$
and $\Sigma_S$ such that $A \models_{\T} a \approx t$.  
\end{itemize}
\end{definition}
\begin{definition} Let $\T$ be a theory with uninterpreted function symbols in a set
$\Sigma_1$. Let $\Sigma_S \subseteq \Sigma_1 \cup C$.
$\T$ has the {\em Beth definability property w.r.t.\ $\Sigma_S$}
($\Sigma_S$-Beth definability), 
if for every conjunction of literals $A$ and every $a \in C$, if $A$ implicitly defines
$a$ w.r.t.\ $\Sigma_S$ and $\T$ then $A$ explicitly
defines $a$ w.r.t.\ $\Sigma_S$ and $\T$.
\end{definition}
In \cite{BruttomessoGR14,CalvaneseGGMR22} it was proved that if a
convex theory $\T$ has the
$\approx$-interpolation property, then it has a version of Beth
definability property which can be regarded as the extension of the 
classical Beth definability property 
\cite{ChangKeisler1990} 
so as to take into account the theory $\T$.   
We give an analogous implication between $\approx$-interpolation
and Beth definability w.r.t.\ a subsignature. 
\begin{thm}
\label{convex-implies-beth}
Let $\T$ be a convex theory with signature $\Pi = (\Sigma_0 \cup
\Sigma_1, {\sf Pred})$, $C$ a set of constants, and 
$\Sigma_S \subseteq \Sigma_1 \cup C$. 
Let $\T'$ be as defined above. 
\begin{enumerate}[(i)]
\item If $\T \cup \T'$ has the $\approx$-interpolation property with 
intersection-sharing,   then $\T$ has the $\Sigma_S$-Beth definability property. 
\item Assume that $\T = \T_0 \cup \K$ where all symbols in the
  signature of $\T_0$ are regarded as interpreted, and $\K$ is a set
  of clauses also containing uninterpreted function symbols in
  $\Sigma_1$. Let $\Theta_{\K}$ be the closure operator defined in
  Example~\ref{ex-shared}.
  If $\T \cup \T'$ has the $\approx$-interpolation property with
  $\Theta_{\K \cup \K'}$-sharing, then $\T$ has the $\Theta_{\K}(\Sigma_S)$-Beth definability property. 
\end{enumerate}
\end{thm}
\noindent {\em Proof:} (i) Assume $a$ is implicitly definable w.r.t.\ $\Sigma_S$, i.e. there
exists a conjunction $A$ of literals such that if $A'$ is obtained by
renaming the symbols in $(\Sigma_1 \cup C) \backslash \Sigma_S$ 
as explained at the beginning of this subsection, we have $A \wedge A' \models_{\T \cup \T'} a \approx a'$. 
Since $\T \cup \T'$ has the $\approx$-interpolation property, there exists a term $t$ using only the
functions and predicate symbols common to $A$ and $A'$ 
such that  $A \wedge A' \models_{\T \cup \T'} a \approx t \wedge t \approx a'$. 
The only function symbols shared by $A$ and $A'$ are the symbols in
$\Sigma_0  \cup \Sigma_S$. We show that $A \models_{\T } a \approx
t$. Let ${\cal A}$ be a model of $\T$ in which $A$ is true. We define
new function symbols $\{ f'_{\cal A} \mid f' \in \Sigma'_r \}$ by
$f'_{\cal A} = f_{\cal A}$, and interpretations $c'_{\cal A} :=
c_{\cal A}$ for all constants not in $\Sigma_S$. The structure  ${\overline
  \A}$ obtained this way is a model of $(\T \cup \T')$ and of $A
\wedge A'$.
Therefore, since $A \wedge A' \models_{\T \cup \T'} a \approx t \wedge
t \approx a'$,  we have    
${\overline {\cal A}} \models a \approx t$, hence ${\cal A} \models a
\approx t$. Therefore, $A \models_{\T} a \approx t$.

\noindent (ii) Assume $a$ is implicitly definable w.r.t.\ $\Theta_\K(\Sigma_S)$, i.e. there
exists a conjunction $A$ of literals such that if $A'$ is obtained
from $A$ and $\T' = \T_0 \cup
\K'$ is obtained from $\T = \T_0 \cup \K$ by
renaming as explained at the beginning of this
subsection,  then $A \wedge A' \models_{\T \cup \T'} a \approx a'$. 
\noindent The function symbols shared by $A$ and $A'$ are the
symbols in $\Sigma_0 \cup \Theta_{\K \cup
  \K'}(\Sigma_S)$, where: 

$\Theta_{\K \cup \K'}(\Sigma_S)  = \bigcup_{f \in \Sigma_S} \{ g \in
\Sigma_1 \cup \Sigma'_1 \mid f \sim^*_{\K \cup \K'}  g \}$

\noindent Note that: 
 
\noindent $\Theta_{\K}(\Sigma_S) = \bigcup_{f \in \Sigma_S} \{ g \in
\Sigma_1 \mid f \sim^*_{\K} g \}$ and 
$\Theta_{\K'}(\Sigma_S) = \bigcup_{f \in \Sigma_S} \{ g \in
\Sigma'_1 \mid f \sim^*_{\K'} g \}$.

\noindent It is easy to see that for
every $f \in \Sigma_1 \backslash \Sigma_S$, $f \in
\Theta_{\K}(\Sigma_S)$ iff $f' = \Theta_{\K'}(\Sigma_S)$, and 
$\Theta_{\K \cup \K'}(\Sigma_S) = \Theta_{\K}(\Sigma_S) \cup
  \Theta_{\K'}(\Sigma_S)$. 

\noindent Since we assumed that $\T \cup \T'$ has the $\approx$-interpolation
property with the notion of $\Theta_{\K \cup \K'}$-sharing, there
exists a term $t$ over the signature $\Sigma_0 \cup \Theta_{\K \cup
  \K'}(\Sigma_S)$ such that $A \wedge  A' \models_{\T \cup \T'} a \approx t ~ \wedge ~ t \approx a$.
The term $t$ might contain primed versions of function symbols. 
We
show that we can find a term ${\overline t}$ containing only terms in $\Theta_\K(\Sigma_S)$
such that $A \models_{\T } a \approx
{\overline t}$. Let ${\cal A}$ be a $\Pi^C$-model of $\T = \T_0 \cup \K$ in which $A$ is
true. For every $f \in \Sigma_1 \backslash
\Theta_\K(\Sigma_S)$ we define  $f'_{\cal A} = f_{\cal A}$, and for
every constant $c$ not in $\Sigma_S$ we define $c'_{\cal A} :=
c_{\cal A}$. The structure  ${\overline \A}$ obtained this way is a
model of $\T_0 \wedge \K$ and of $A$ (because $\A$ was a model of $\T_0 \cup
\K$ and of $A$), a model of $\T_0 \cup \K'$ and of $A'$ (because of the definition of the
primed symbols), so  
${\overline {\cal A}} \models a \approx t$. Since for $f \in \Sigma_1
\backslash \Theta_\K(\Sigma_S)$ the functions $f$
and $f'$ are defined in the same way in ${\overline {\cal A}}$, and
the primed and unprimed versions of constants in $C \backslash
\Sigma_S$ are interpreted the same way in ${\overline {\cal A}}$, we can
replace in $t$ all primed function symbols with the non-primed
versions, and its value does not change in ${\overline {\cal A}}$. 
Let ${\overline t}$ be the term obtained this way. The value of
${\overline t}$ in ${\cal A}$ is equal to the value of $t$ in
${\overline {\cal A}}$, i.e.\ is equal to the value of $a$ in ${\cal
  A}$. Hence, ${\cal A} \models a
\approx {\overline t}$.
\QED

\section{Local Theory Extensions} 
\label{local}

Let $\Pi_0 {=} (\Sigma_0, {\sf Pred})$ be a signature, and ${\cal T}_0$ be a 
``base'' theory with signature $\Pi_0$. 
We consider 
extensions $\T := {\cal T}_0 \cup \K$
of ${\cal T}_0$ with new function symbols $\Sigma_1$
({\em extension functions}) whose properties are axiomatized using 
a set $\K$ of (universally closed) clauses 
in the extended signature $\Pi = (\Sigma_0 \cup \Sigma_1, {\sf Pred})$, 
which contain function symbols in $\Sigma_1$. 

\medskip
\noindent 
If $G$ is a finite set of ground $\Pi^C$-clauses, where $C$ is an
additional set of constants, and $\K$ a set of $\Pi$-clauses, we 
will denote by ${\sf st}({\cal K}, G)$ the set of all 
ground terms which occur in $G$ or ${\cal K}$, and by 
${\sf est}({\cal K}, G)$ the set of all extension ground terms, i.e.\ 
terms starting with a function in $\Sigma_1$
which occur in $G$ or ${\cal K}$. 

\medskip
\noindent  In this paper we regard every finite set $G$
of ground clauses as the ground formula $\bigwedge_{C \in G} C$. 
%
If $T$ is a set of ground terms in the signature  $\Pi^C$, 
we denote by $\K[T]$ the set of all instances of $\K$ in which the terms 
starting with a function symbol in $\Sigma_1$ are in $T$. 
Let $\Psi$ be a map associating with 
every finite set $T$ of ground terms a finite set $\Psi(T)$ of ground
terms containing $T$. 
For any set $G$ of ground $\Pi^C$-clauses we write 
$\K[\Psi_{\cal K}(G)]$ for $\K[\Psi({\sf est}({\cal K}, G))]$.
We define the following condition: 
 
\medskip
\noindent \begin{tabular}{ll}
${\sf (Loc}_f^\Psi)$~ &  For every finite set $G$ of ground clauses in
$\Pi^C$ it holds that\\
&  $\T_0 \cup {\cal K} \cup G \models \bot$ if and only if $\T_0
\cup \K[\Psi_{\cal K}(G)] \cup G$ is unsatisfiable. 
\end{tabular}

\medskip
\noindent Extensions satisfying condition ${\sf (Loc}_f^\Psi)$ are called
{\em $\Psi$-local}. 
If $\Psi$ is the identity 
we obtain the notion of {\em local theory  
extensions} \cite{sofronie-cade-05}; if in addition $\T_0$ is
the theory of pure equality we obtain the notion of 
{\em local theories}  \cite{McAllester93,Ganzinger-01-lics}.

\subsection{Hierarchical reasoning in local theory extensions}
Consider a $\Psi$-local theory extension 
${\cal T}_0 \subseteq {\cal T}_0 \cup {\cal K}$.
Condition $({\sf Loc}_f^{\Psi})$ requires that for every finite set $G$ of ground 
$\Pi^C$-clauses, ${\cal T}_0 \cup {\cal K} \cup G \models \perp$ iff 
${\cal T}_0 \cup {\cal K}[\Psi_{\cal K}(G)] \cup G \models \perp$.
In all clauses in ${\cal K}[\Psi_{\cal K}(G)] \cup G$ the function 
symbols in $\Sigma_1$ only have ground terms as arguments, so  
${\cal K}[\Psi_{\cal K}(G)] {\cup} G$ can be flattened 
and purified
by introducing, in a bottom-up manner, new  
constants $c_t \in C$ for subterms $t {=} f(c_1, \dots, c_n)$ where $f {\in}
\Sigma_1$ and $c_i$ are constants, together with 
definitions $c_t {=} f(c_1, \dots, c_n)$. 
We thus obtain a set of clauses ${\cal K}_0 {\cup} G_0 {\cup} {\sf Def}$, 
where ${\cal K}_0$ and $G_0$ do
not contain $\Sigma_1$-function symbols and ${\sf Def}$ contains clauses of the form 
$c {=} f(c_1, \dots, c_n)$, where $f {\in} \Sigma_1$, $c, c_1, \dots,
c_n$ are constants.
\begin{thm}[\cite{sofronie-cade-05,ihlemann-jacobs-sofronie-tacas08,sofronie-ihlemann-10}]
Let ${\cal K}$ be a set of clauses. 
Assume that 
${\cal T}_0 \subseteq {\cal T}_0 \cup {\cal K}$ is a 
$\Psi$-local theory extension. 
For any finite set $G$ of flat ground clauses (with no nestings of
extension functions), 
let ${\cal K}_0 \cup G_0 \cup {\sf Def}$ 
be obtained from ${\cal K}[\Psi_{\cal K}(G)] \cup G$ by flattening and purification, 
as explained above. 
Then the following are equivalent to ${\cal T}_0 \cup {\cal K} \cup G \models \perp$: 
\begin{itemize}
\item[(i)] ${\cal T}_0 {\cup} {\cal K}[\Psi_{\cal K}(G)] {\cup} G \models
  \perp.$ 
\item[(ii)] ${\cal T}_0 \cup {\cal K}_0 \cup G_0 \cup {\sf Con}_0 \models \perp,$ where 
${\sf Con}_0  {=}$ \small{$\displaystyle{\{ {\bigwedge}_{i =
      1}^n \!\! c_i
    {\approx} d_i {\rightarrow} c {\approx} d \, {\mid}
\begin{array}{@{}l@{}}
f(c_1, \dots, c_n) {\approx} c {\in} {\sf Def}\\
f(d_1, \dots, d_n) {\approx} d {\in} {\sf Def} 
\end{array} \}}.$} 
\end{itemize}
\label{lemma-rel-transl}
\end{thm} 
The locality of a theory extension 
can be recognized by proving embeddability of partial models into 
total models 
\cite{sofronie-cade-05,sofronie-ihlemann-ismvl-07,ihlemann-jacobs-sofronie-tacas08}. 
In \cite{sofronie-ihlemann-10} we showed that for extensions with sets
of flat and linear clauses $\Psi$-locality can be
checked by checking whether an embeddability condition of partial into
total models holds. In \cite{sofronie-fuin-2017} we mention (without
proof) that the proof
in \cite{sofronie-ihlemann-10} can be extended to situations in which 
the clauses in $\K$ are not linear. 
A full proof of this result is given in Section~\ref{recogn-non-lin}.  
For presenting the proof we 
need to introduce notions such as partial structures and truth in 
partial structures.

\subsection{Partial structures, weak validity}

\noindent We first introduce notions such as partial structures, truth in 
partial structures, and weak partial models for clauses resp.\ for
sets of clauses. 
\begin{definition}
A {\em partial $\Pi$-structure} is a structure $\A = (A, \{ f_{\A}
\}_{f \in \Sigma}, \{ p_{\A} \}_{p \in {\sf Pred}})$,  
where $A$ is a non-empty set and for every $f \in \Sigma$ with arity $n$, 
$f_{\A}$ is a partial function from $A^n$ to $A$ and for every $p \in
{\sf Pred}$ with arity $m$, $p_{\cal A} \subseteq A^m$.
A {\em (total) $\Pi$-structure} is a partial $\Pi$-structure where 
all functions $f_{\A}$ are total.
\end{definition} 
\noindent 
The notion of evaluating a term $t$ w.r.t.\ a variable assignment 
$\beta : X \rightarrow \A$ for its variables in a partial structure $\A$ 
is the same as for a total structure, except that this evaluation is undefined
if $t = f(t_1, \dots, t_n)$ and either one of $\beta(t_i)$ is undefined, or 
else $(\beta(t_1), \dots, \beta(t_n))$ is not in the domain of
definition of 
$f_{\A}$. 

\begin{definition}
Let $D$ be a clause. We say that $(\A, \beta)$ is a weak partial model
of $D$ (notation: $(\A, \beta) \models_w D$) if either at least one term in $D$ is not
defined in $\beta$ or all terms are defined and at least one literal
$L$ of $D$ holds in $(\A, \beta)$. 
\end{definition}

\begin{definition}
A total map $h : A \rightarrow B$ between partial $\Pi$-structures $\A$ and $\B$
is called a {\em weak $\Pi$-ho\-mo\-mor\-phism} if whenever 
$f_A(a_1, \dots, a_n)$ is defined in $A$, also 
$f_B(h(a_1), \dots, h(a_n))$ is defined in $B$ and 
$h(f_A(a_1,\dots,a_n)) = f_B(h(a_1),\dots,h(a_n))$. 
A partial structure $\A$ weakly embeds into a (total) structure $\B$ 
if there exists 
an injective weak $\Pi$-homomorphism $h$ from $\A$ to $\B$ such that 
for every $R \in {\sf Pred}$ and for every $a, b \in \A$ we have: 
$a R b$ iff $h(a) R h(b)$.
\end{definition}

\subsection{Flat and linear clauses; Flattening and purification} 

We define flatness for non-ground clauses
and for ground clauses.
\begin{definition}[Flatness and linearity for non-ground clauses]
A non-ground clause is $\Sigma_1$-{\em flat} 
if function symbols (including constants) do not occur 
as arguments of function symbols in $\Sigma_1$.
A $\Sigma_1$-flat non-ground clause is called $\Sigma_1$-{\em linear} 
if whenever a variable occurs in two terms in the clause 
which start with function symbols in $\Sigma_1$, 
the two terms are identical, and if 
no term which starts with a function symbol 
in $\Sigma_1$ contains two occurrences of the same variable.
\end{definition}
\begin{definition}[Flatness and linearity for ground clauses]
We say that a ground clause is $\Sigma_1$-{\em flat} if only constants 
appear as arguments of 
function symbols in $\Sigma_1$.
A $\Sigma_1$-flat ground clause is $\Sigma_1$-{\em linear} 
if whenever a 
constant occurs in two terms in the clause 
starting with function symbols in $\Sigma_1$, 
the two terms
are identical, and if no term starting with a function symbol 
in $\Sigma_1$ contains two occurrences of the same constant.
\end{definition}
Any set $G$ of ground clauses in a signature $\Sigma$ 
containing $\Sigma_1$
can be transformed into a set $G_{{\sf flin}(\Sigma_1)}$ 
of ground clauses in which subterms starting 
with function symbols in $\Sigma_1$ are flat and linear. This can be 
done by introducing, in a bottom-up manner,
new constants for subterms occurring below functions in $\Sigma_1$, 
and adding the corresponding definitions to the set of clauses.

A set $G$ of ground clauses can be transformed into a {\em purified}\/
set of clauses  $G_{{\sf sep}(\Sigma_1)}$
(i.e.\ the function symbols in $\Sigma_1$ are separated from 
the other symbols)  
by introducing, in a bottom-up manner,
new  constants $c_t$ for
subterms $t = f(g_1, \dots, g_n)$ with $f \in \Sigma_1$, $g_i$ ground 
$\Sigma_0 \cup \Sigma_c$-terms (where $\Sigma_c$ is a set of constants 
which contains the constants introduced by flattening), together with 
corresponding definitions $c_t \approx t$.
These transformations preserves satisfiability and unsatisfiability
with respect to total algebras, and also with respect to 
partial algebras in which all ground subterms 
which are flattened are defined.

\subsection{Recognizing locality and $\Psi$-locality}
\label{recogn-non-lin}
We now give a semantic criterion for $\Psi$-locality.

\begin{definition}
With the above notations, let $\Psi$ be a map associating with 
${\cal K}$ and a set of ground terms $T$ 
a set ${\Psi}_{\cal K}(T)$ of ground terms. 
We call ${\Psi}_{\cal K}$ a \emph{term closure operator} if the following 
holds for all sets of ground terms $T, T'$: 
\begin{itemize}
\vspace{-2mm}
\item[(i)] $\mathrm{est}({\cal K}, T) \subseteq {\Psi}_{\cal K}(T)$,
\item[(ii)] $T \subseteq T' \Rightarrow \Psi_{\cal K}(T) \subseteq 
  \Psi_{\cal K}(T')$,
\item[(iii)] $\Psi_{\cal K}(\Psi_{\cal K}(T)) \subseteq \Psi_{\cal K}(T)$,
\item[(iv)] for 
  any map $h: C \rightarrow C$, $\bar{h}(\Psi_{\cal K}(T)) =
  \Psi_{\cal K}(\bar{h}(T))$,
  where $\bar{h}$ is the canonical extension of $h$ to extension 
  ground terms. 
\vspace{-2mm}
\end{itemize}
\end{definition}
We will use the following notation: If $A$ is a set (for instance the 
support of a $\Pi$-structure), we denote by $\Pi^A$ the extension of 
$\Pi$ with the elements in $A$, seen 
  as additional constants. Let $\T = \T_0 \cup \K$ be a theory 
  extension with set of clauses $\K$ and let $\Psi_\K$ be a closure 
  operator.  

\smallskip 
\noindent 
In what follows, 
${\sf PMod^{\Psi}_w}({\Sigma_1}, {\mathcal T})$ denotes 
the class of all partial structures $\A = (A, \{ f_{\A} \}_{f \in 
  \Sigma_0 \cup \Sigma_1}, \{ p_{\A} \}_{p \in {\sf Pred}})$ 
in which  the $\Sigma_0$-functions are total, and $\A_{|\Pi_0}$ is a 
model of ${\cal T}_0$, the $\Sigma_1$-functions are partial and 
$\A$ is a  weak partial model of all the clauses in ${\cal K}$,  
and in which  
the set of $\Pi^A$ terms 

\smallskip
${\sf Def}(A) = \{ f(a_1, \dots, a_n) \mid
a_1, \dots, a_n \in A, f_{\A}(a_1, \dots,
a_n) \text{ defined} \}$

\smallskip
\noindent is closed under $\Psi_{\cal K}$.

\medskip 
\noindent For extensions 
${\mathcal T}_0 \subseteq  {\mathcal T} = {\mathcal T}_0 \cup {\mathcal K}$,
where ${\mathcal K}$ is a set of clauses, 
we consider the 
condition:

\medskip 
\noindent \begin{tabular}{@{}ll}
${\sf (Emb^{\Psi}_w)}$  & Every 
$A \in {\sf PMod^{\Psi}_w}({\Sigma_1}, {\mathcal T})$ 
weakly embeds into a total model of ${\mathcal T}$. 
\end{tabular}

\begin{thm} 
Let ${\mathcal K}$ be a set of $\Sigma_1$-flat clauses, and 
$\Psi_\K$ be a term closure operator, which satisfies conditions
(i)--(iv) above and also satisfies the additional condition that for 
every set $T$ of ground terms and  
for every clause  $D$ in ${\mathcal K}$, if a variable occurs in two
terms in $D$ then 
either the two terms are identical, or the variable occurs 
below two different unary function symbols $f$ and $g$ 
and, for every constant $c$, $f(c)$ is in $\Psi_\K(T)$ iff  
$g(c)$ is in $\Psi_\K(T)$. 
If the extension 
${\mathcal T}_0 \subseteq {\mathcal T} = {\cal T}_0 \cup {\cal K}$ 
satisfies ${\sf (Emb^{\Psi}_w)}$ 
then the extension satisfies ${\sf (Loc^{\Psi})}$.
\label{rel-loc-embedding}
\end{thm}
{\em Proof:}  Assume that ${\mathcal T}_0 \cup {\mathcal K}$ is not a
$\Psi$-local extension of ${\mathcal T}_0$. Then there exists 
a set $G$ of ground clauses (with additional constants) 
such that 
${\mathcal T}_0 \cup {\mathcal K} \cup G \models \perp$ but 
${\mathcal T}_0 \cup {\mathcal K}[\Psi_{\cal K}(G)] \cup G$ 
has a model ${\cal B}$.  
We assume w.l.o.g.\ that $G = G_0 \cup G_1$, 
where $G_0$ contains no function symbols in 
$\Sigma_1$ and $G_1$ consists of ground unit clauses of the form
$f(c_1, \dots, c_n) \approx c,$
where 
$c_i, c$ are constants in $\Sigma_0 \cup C$
and $f \in \Sigma_1$.

\noindent We construct another structure,  $\A$, having the same
support $U_\A$ as $\B$, 
which inherits all relations in ${\sf Pred}$ and 
all maps in $\Sigma_0 \cup C$ from $\B$, but on which  
the domains of definition of the $\Sigma_1$-functions are restricted 
as follows: for every $f \in \Sigma_1$, 
$f_{\A}(a_1, \dots, a_n)$ is defined if and only if 
there exist constants $c^1, \dots, c^n$ such that 
$f(c^1, \dots, c^n)$ is in $\Psi_{\cal K}(G)$ and 
$a_i = c^i_{\B}$ for all $i \in \{ 1, \dots, n \}$. 
In this case we define $f_{\A}(a_1, \dots, a_n) := f_{\B}(c^1_{\B}, \dots, c^n_{\B})$.
The reduct of $\A$ to $(\Sigma_0 \cup C, {\sf Pred})$ 
coincides with that of $\B$. 
Thus, $\A$ is a model of ${\mathcal T}_0 \cup G_0$. 
By the way the operations in $\Sigma_1$ are defined in $\A$ it is 
clear that $\A$ satisfies $G_1$, so $\A$ satisfies $G$. 

\smallskip
\noindent We show that $\A \models_w {\mathcal K}$.
Let $D$ be a clause in ${\mathcal K}$. 
If $D$ is ground then all its terms are defined (and all terms
starting with an extension function are contained in $\Psi_\K(G)$), i.e.\ 
$D \in {\mathcal K}[\Psi_\K(G)]$, so $D$ 
is true in $\B$, hence it is also true in $\A$.

Now consider the case in which $D$ is not ground. 
Let $\beta : X \rightarrow \A$ be an arbitrary
valuation. Again, if there is a term $t$ in $D$ such that $\beta(t)$ is undefined, we
immediately have that $(\A, \beta)$ weakly satisfies $D$. 
So let us suppose that for all terms $t$ occurring in $D$, $\beta(t)$
is defined. We associate with $\beta$ a substitution $\sigma$ as follows: 
Let $x$ be a variable. We have the following possibilities: 

\smallskip
\noindent {\em Case 1:} $x$ does not occur below any extension
function. This case is unproblematic. We can define $\sigma(x)$ arbitrarily.

\smallskip
\noindent {\em Case 2:}  $x$ occurs in a unique term $t = f(...x...y...)$ 
(which may occur more than once in the clause $D$). From the fact that $\beta(t)$ is 
defined, we know that
there are ground terms (in fact constants) which we will denote by $t_x, t_y, \dots$ such
that $\beta(x) =(t_x)_{\B}, \beta(y) =(t_y)_{\B}, \dots$, $\beta(t) =
f_A(...(t_x)_{\B} \dots (t_y)_{\B} \dots)$, and $f(\dots, t_x, \dots, t_y,
\dots) \in \Psi_{\cal K}(G)$. 
We can define $\sigma(x) = t_x$. 

\smallskip
\noindent {\em Case 3:}  $x$ occurs in two terms of the form $f(x),
g(x)$ in the clause $D$. By assumption, $\Psi_{\K}$ has the property that  for every constant
$c$, $f(c) \in \Psi_{\cal K}(G)$ iff $g(c) \in \Psi_{\cal K}(G)$. 
From the fact that $\beta(f(x))$ and $\beta(g(x))$ are 
defined, we know that
there are ground terms which we will denote by $t_x, s_x$ such
that $\beta(x) =(t_x)_{\B} =(s_x)_{\B}$, $\beta(f(x)) = f(t_x)_{\B}$
and $\beta(g(x)) = g(s_x)_{\B}$, and 
$f(t_x), g(s_x) \in \Psi_{\cal K}(G)$. 
Since all terms in $\Psi_{\cal K}(G)$ are flat, $t_x$ and $s_x$ are
constants. Assume that $t_x = c$. 
We know that if $f(c) \in \Psi_{\cal K}(G)$ then $g(c) \in \Psi_{\cal  K}(G)$ and if 
$g(c) \in \Psi_{\cal  K}(G)$ then $f(c) \in \Psi_{\cal  K}(G)$, 
so we can choose $s_x = t_x = c$. Also in this case 
we can define $\sigma(x) = t_x = c$. 

\smallskip
\noindent 
Thus, we can construct a substitution $\sigma$ with 
$\sigma(D) \in {\mathcal K}[\Psi_{\K}(G)]$ and $\beta \circ \sigma = \beta$. 
As $(\B, \beta) \models \sigma(D)$ we can infer $(\A, \beta) \models_w D$.
We now show that $${\sf Def}(\A) = \{ f(a_1, \dots, a_n) \mid a_1,
\dots, a_n \in U_{\A}, f_{\A}(a_1, \dots, a_n) \text{ defined} \}$$ 
is closed under $\Psi_{\cal K}$. 
By definition, $f(a_1, \dots, a_n) \in {\sf Def}(\A)$ iff  there exist 
$\text{ constants } c_1, \dots, c_n$ with ${c_i}_{\A} = a_i$ for all $i$ and $f(c_1, \dots, c_n) \in \Psi_{\cal K}(G)$. Thus, 
$$\begin{array}{rll}
{\sf Def}(\A) & =  \{ f(a_1, \dots, a_n) \mid a_1, \dots, a_n \in U_{\A}, f_{\A}(a_1, \dots, a_n) \text{ defined} \} & \\
      & =  \{ f({c_1}_{\A}, \dots, {c_n}_{\A}) \mid c_i \text{ constants with } f(c_1, \dots, c_n) \in \Psi_{\cal K}(G) \} & \\
      & =  {\overline h}(\Psi_{\cal K}(G)) & \!\!\!\!\!\!\!\!\!\!\!\!\!\!\!\!\!\!\!\!\!\!\!\!\!\!\!\!\!\!\!\!\!\!\!\!\!\!\!\!\!\!\!\!\!\!\!\!\!\!\!\!\!\!\!\!\!\!\!\!\!\!\!\text{ where } h(c_i) = a_i \text{ for all } i \\
\Psi_{\cal K}({\sf Def}(\A)) & =  \Psi_{\cal K}({\overline h}(\Psi_{\cal K}(G)))  = {\overline h}(\Psi_{\cal K}(\Psi_{\cal K}(G))) & \!\!\!\!\!\!\!\!\!\!\!\!\!\!\!\!\!\!\!\!\!\!\!\!\!\!\!\!\!\!\!\!\!\!\!\!\!\!\!\!\!\!\!\!\!\!\!\!\!\!\!\!\!\!\!\!\!\!\!\!\!\!\!\text{ by property (iv) of } \Psi \\
& \subseteq {\overline h}(\Psi_{\cal K}(G)) = {\sf Def}(\A) & \!\!\!\!\!\!\!\!\!\!\!\!\!\!\!\!\!\!\!\!\!\!\!\!\!\!\!\!\!\!\!\!\!\!\!\!\!\!\!\!\!\!\!\!\!\!\!\!\!\!\!\!\!\!\!\!\!\!\!\!\!\!\!\text{ by property (iii) of } \Psi\\

\end{array}$$
As $\A \models_w {\mathcal K}$, 
$\A$ weakly embeds into a total structure ${\overline \A}$ satisfying 
${\mathcal T}_0 \cup {\mathcal K}$. 
But then ${\overline \A} \models G$, 
so ${\overline \A} \models {\mathcal T}_0 \cup {\mathcal K} \cup G$, 
which is a contradiction. \QED

\section{$R$-interpolation in local theory extensions}
\label{p-int}

In \cite{Sofronie-lmcs} we considered convex and $P$-interpolating
theories $\T_0$ with signature $\Pi_0 =
(\Sigma_0, {\sf Pred})$  (where $P
\subseteq {\sf Pred}$).  We studied $\Psi$-local extensions 
$\T = \T_0 \cup \K$ of a theory $\T_0$ with new function symbols in
a set $\Sigma_1$ axiomatized by a set $\K$ of clauses, with the property that all
clauses in $\K$ are of the form: 
\begin{eqnarray}
\left\{ \begin{array}{l} x_1 \, R_1 \, s_1 \wedge \dots 
\wedge x_n \, R_n \, s_n \rightarrow 
f(x_1, \dots, x_n) \, R \, g(y_1, \dots, y_n) \\ 
x_1 \, R_1 \, y_1 \wedge \dots 
\wedge x_n \, R_n \, y_n \rightarrow 
f(x_1, \dots, x_n) \, R \, f(y_1, \dots, y_n) \end{array} \right.
\label{general-form} 
\end{eqnarray}
\begin{quote}
where $n \geq 1$, $x_1, \dots, x_n, y_1, \dots, y_n$ are variables,
$f, g \in \Sigma_1$, $R_1, \dots, R_n, R$ 
are binary relations with $R_1, \dots, R_n \in P$ and $R$ transitive, 
and for each $i \in \{ 1, \dots, n \}$, the term $s_i$ is either a variable 
among the arguments of $g$, or a term of the form $h_i(z_1, \dots, z_k)$, 
where $h_i \in \Sigma_1$ and all the arguments of $h_i$ are 
variables occurring  among the arguments of $g$.  
\end{quote}
\begin{example}
A set $\K$ of axioms containing clauses of the form: 
$$\left\{ \begin{array}{rl}
x_1 \leq h(y_1) & \rightarrow f(x_1) \leq g(y_1) \\
x_1 \leq y_1 & \rightarrow f(x_1) \leq f(y_1)  
\end{array}
\right.$$
satisfies the conditions above: $n = 1$, $R_1 = R = \,
\leq$, $s_1 = h(y_1)$, $f, g, h \in \Sigma_1$.
\end{example}
We make the following assumptions (which were also made in
\cite{Sofronie-lmcs}) on the theory $\T_0$ and its theory extension
$\T_0 \cup \K$: 
\begin{description}
\item[A1:] $\T_0$ is convex and has the $P$-interpolation property.
\item[A2:] Satisfiability of ground clauses w.r.t.\ $\T_0$ is
  decidable.
\item[A3:] All
clauses in $\K$ are of the form~(\ref{general-form}). 
\item[A4:] $\T_0 \subseteq \T_0 \cup \K$ is a $\Psi$-local extension.
\end{description}
In \cite{Sofronie-lmcs}, we proved that if $\T_0$ allows ground interpolation,  
then $\T$ allows ground interpolation, and that the interpolants can  
be computed in a hierarchical way, using a method for ground  
interpolation in $\T_0$. 
 
\medskip
\noindent We show that under the conditions above, the property of {\em $P$-interpolation}
can be transferred from the theory $\T_0$ to the extension $\T 
= \T_0 \cup \K$ of $\T_0$.
The function symbols in the signature of $\T_0$ are considered to be
interpreted, and will always be considered to be shared. For the function 
symbols in the signature $\Sigma_1$ -- considered to be
``quasi''-interpreted -- we use the notion of $\Theta_{\K}$-sharing
introduced in Section~\ref{prelim}.

\medskip
\noindent In order to show that $\T$ has the $P$-interpolation
property, we need to prove that if $A$, $B$ are conjunctions of atoms
and $A({\overline c}, {\overline a}_1, a) \wedge B({\overline c},
{\overline b}_1, b) \models_{\T} a R b$, 
where $R \in P$, then there exists a term $t$ containing only the
constants common to $A$ and $B$ and only function symbols which are
{\em $\Theta_\K$-shared} by $A$ and $B$, such that $A({\overline c}, \overline{a}_1, a) \wedge B({\overline c},
\overline{b}_1,  b) \models_{\T} a R t ~\wedge~  t R b$.

\medskip
\noindent 
Note that 
$$A({\overline c}, \overline{a}_1, a) \wedge B({\overline c},
\overline{b}_1, b) \models_{\T} a R b \text{ iff } A({\overline c}, \overline{a}_1, a) \wedge B({\overline c},
\overline{b}_1,  b) \wedge \neg (a R b) \models_{\T} \bot.$$ 
By Assumption {\bf A4}, $\T_0 \cup \K$ is a $\Psi$-local extension of
$\T_0$. We can purify and flatten this
conjunction and obtain a conjunction of unit clauses $A_0 \wedge B_0 \wedge {\sf
  Def} \wedge \neg (a R b)$, where ${\sf Def}$ is a set of definitions of
newly introduced constants.  Let $T$ be the extension terms in ${\sf Def}$. We introduce
new constants and definitions also for all extension terms in $\Psi(T)$. This new
set of definitions can be written as a conjunction $D_A \wedge D_B$ of its $A$-part and its
$B$-part. \label{page}
By the $\Psi$-locality of the extension $\T_0 \subseteq \T_0 \cup \K$ and 
Theorem~\ref{lemma-rel-transl}, 
$$A_0 \wedge B_0 \wedge {\sf Def} \wedge \neg (a R b) 
\models_{{\cal T}} \perp \text{ iff } 
{\cal K}_0 \wedge A_0 \wedge B_0 \wedge {\sf Con}[D_A \wedge D_B]_0 
\wedge \neg (a R b) \models_{{\cal T}_0} 
\perp,$$
where
${\cal K}_0$ is obtained from 
${\cal K}[D_A \wedge D_B]$ by replacing the $\Sigma_1$-terms with the 
corresponding constants contained in the definitions $D_A \wedge D_B$ and 
 
\noindent 
${\sf Con}[D_A \wedge D_B]_0 = \displaystyle{ \bigwedge  \Big\{ \bigwedge_{i = 1}^n} c_i \approx d_i   \rightarrow c \approx d 
\mid \begin{array}{l}
          f(c_1, \dots, c_n) \approx c \in D_A \cup D_B, \\
          f(d_1, \dots, d_n) \approx d \in D_A \cup D_B 
\end{array} \Big\}.$

\smallskip
\noindent 
${\sf Con}[D_A \wedge D_B]_0 = {\sf Con}^A_0 \wedge {\sf Con}^B_0 \wedge {\sf Con}_{\sf mix}$ and 
${\cal K}_0 = {\cal K}^A_0 \wedge {\cal K}^B_0 \wedge {\cal K}_{\sf mix}$,
where 
\begin{itemize}
\item ${\sf Con}^A_0, {\cal K}^A_0$ only  contain
extension functions and 
constants which occur in $A$, 
\item ${\sf Con}^B_0, {\cal K}^B_0$ only  contain
extension functions and constants which occur in $B$, 
\item ${\sf Con}_{\sf mix}$, 
${\cal K}_{\sf mix}$ contain mixed clauses with constants occurring in 
$A$ and in $B$.  
\end{itemize}
Our goal is to separate ${\sf Con}_{\sf mix}$ and 
${\cal K}_{\sf mix}$ into an $A$-part and a $B$-part, which
would allow us to use the $P$-interpolation property of theory $\T_0$.
\begin{prop}
Assume that $\T_0$ is convex and $P$-interpolating and the
intermediate terms can be effectively computed. Assume that ground
satisfiability w.r.t.\ $\T_0$ is decidable. 

\noindent Let ${\cal H}$ be a set of Horn clauses 
$(\bigwedge_{i = 1}^n c_i R_i d_i) \rightarrow c R_0 d$ 
in the signature $\Pi_0^C$ (with $R_0$ transitive and $R_i \in P$) 
which are instances of flattened and purified clauses of 
type~(\ref{general-form}) and of congruence axioms. 
Let ${\cal H}_{\sf mix}$ be the mixed clauses in ${\cal H}$: 

\smallskip
\noindent 
$\begin{array}{@{}lll}
{\cal H}_{\sf mix} & = & \{ \bigwedge_{i = 1}^n c_i R_i d_i \rightarrow c R_0 d \in {\cal H} \mid c_i, c \text{ constants in } A,  d_i, d \text{ constants in } B \} \cup \\
& & \{ \bigwedge_{i = 1}^n c_i R_i d_i \rightarrow c R_0 d \in {\cal H} \mid c_i, c \text{ constants in } B,  d_i, d \text{ constants in } A \} \\[0.5ex]
\end{array}$

\smallskip
\noindent Let $A_0$ and $B_0$ be conjunctions of ground literals
in the signature $\Pi_0^C$ such that 
$A_0 \wedge B_0 \wedge {\cal H} \wedge \neg (a R b) \models_{{\cal T}_0} \perp$. 
Then ${\cal H}$ can be separated into an $A$- and a $B$-part 
by replacing the set ${\cal H}_{\sf mix}$ of mixed clauses with a
separated set of formulae ${\cal H}_{\sf sep}$: 
\begin{enumerate}[(i)]
\item There exists a set $T$ of $(\Sigma_0 \cup C)$-terms containing 
only constants common to $A_0$ and $B_0$ such that 
$A_0 \wedge B_0 \wedge ({\cal H} \backslash {\cal H}_{\sf mix}) 
\wedge {\cal H}_{\sf sep} \wedge \neg (a R b) \models_{{\cal T}_0} \perp$, where 

\medskip
\noindent 
$\begin{array}{@{}l@{}l@{}l}
{\cal H}_{\sf sep} & = & \{ (\bigwedge_{i = 1}^n c_i R_i t_i \rightarrow c R c_{f(t_1, \dots, t_n)}) \wedge (\bigwedge_{i = 1}^n t_i R_i d_i \rightarrow c_{f(t_1, \dots, t_n)} R d)  \mid  \\
& & \bigwedge_{i = 1}^n c_i R_i d_i \rightarrow c R d  \,{\in}\, {\cal H}_{\sf mix}, d_i {\approx} s_i(e_1, \dots, e_n), d {\approx} g(e_1, \dots, e_n) {\in} D_B,  \\
& & c {\approx} f(c_1, \dots, c_n) {\in} D_A \text{ or vice versa } \} = {\cal H}_{\sf sep}^A \wedge {\cal H}_{\sf sep}^B 
\end{array}$

\medskip 
\noindent 
and $c_{f(t_1, \dots, t_n)}$ are new constants in $\Sigma_c$ 
(considered to be common) introduced for the
corresponding terms $f(t_1, \dots, t_n)$, where for $i \in \{ 1,
\dots, n \}$,  $t_i$ separates the atom $c_i
R_i d_i$, which is entailed by the already deduced atoms. 

\medskip
\item $A_0 \wedge B_0 \wedge ({\cal H} \backslash {\cal H}_{\sf mix}) 
\wedge {\cal H}_{\sf sep} \wedge \neg (a R b)$ is logically equivalent with respect to ${\cal T}_0$ with 
the following separated  conjunction of ground literals: 

\medskip
\noindent 
$\begin{array}{@{}l@{}l@{}l}
{\overline A}_0 \wedge {\overline B}_0 \wedge & \neg (a R b) & =   
A_0 \,\wedge\, B_0 \wedge \neg (a R b) \wedge \bigwedge \{ c R d
\mid \Gamma {\rightarrow} c R d \in {\cal H} \backslash {\cal H}_{\sf
  mix} \} \wedge \\
&  &   
\!\!\!\!\! \!\!\!\!\! \!\!\!\!\! \bigwedge \{  c R c_{f({\overline t})} \wedge c_{f({\overline t})} R d \mid 
(\Gamma \rightarrow c R c_{f({\overline t})}) \wedge (\Gamma \rightarrow c_{f({\overline t })} R d) \in {\cal H}_{\sf sep} \}.
\end{array}$
\end{enumerate}
\label{sep}
\end{prop}
{\em Proof:} The proof is similar to that of Proposition~5.7 in
\cite{Sofronie-lmcs}. (i) and (ii) are proved simultaneously by
induction on the number of 
clauses in ${\mathcal H}$. If ${\mathcal H} = \emptyset$, it is
already separated into an $A$ and a $B$ part so we are
done. 
Assume that ${\mathcal H}$ contains at least one clause, and that for 
every ${\mathcal H}'$ with fewer clauses and all conjunctions of 
literals $A'_0, B'_0$ with $A'_0 \wedge B'_0 \wedge {\mathcal H}' \wedge \neg a R b
\models_{{\mathcal T}_0}  \perp$, (i) and (ii) hold. 

\medskip
\noindent 
Let ${\mathcal D}$ be the set of all atoms $c_i R_i d_i$
occurring in premises of clauses in ${\mathcal H}$. 
As every model of $A_0 \wedge B_0 \wedge
\bigwedge_{(c R' d) \in {\mathcal D}} \neg (c R' d) \wedge \neg a R b$
is also a model for
${\mathcal H} \wedge A_0 \wedge B_0 \wedge \neg a R b$ and 
since ${\mathcal H} \wedge A_0 \wedge B_0 \wedge \neg a R b
\models_{{\mathcal T}_0} \perp$,  
it follows that 
$A_0 \wedge B_0 \wedge
\bigwedge_{(c R' d) \in {\mathcal D}} \neg (c R' d) \wedge \neg (a R b)
\models_{{\mathcal T}_0} \perp$.  

\noindent Let  $(A_0 \wedge B_0)^+$ be the conjunction 
of all atoms in $A_0 \wedge B_0$, and $(A_0 \wedge B_0)^-$ be the set of all 
negative literals in $A_0 \wedge B_0$. Then 
$$(A_0 \wedge B_0)^+ \models_{{\mathcal T}_0} \bigvee_{(c R d) \in
  {\mathcal D}} (c R d) \vee \!\!\!\!\!\! \bigvee_{\neg L \in (A_0 \wedge B_0)^-}
\!\!\!\!\!\! L
~~\vee a R b.$$ 
We know that ${\mathcal T}_0$ is convex with respect to ${\sf Pred}$. 
Moreover, $(A_0 \wedge B_0)^+$ is a
conjunction of positive literals. Therefore, either 
\begin{itemize}
\item[(a)] $(A_0 \wedge B_0)^+ \models a R b$, or 
\item[(b)] $(A_0 \wedge B_0)^+ \models L$ for some 
$L \in (A_0 \wedge B_0)^-$ (then 
$A_0 \wedge B_0$ is unsatisfiable and hence 
entails any atom $c_i R_i d_i$), or  
\item[(c)] there exists $(c_1 R_1 d_1) \in {\mathcal D}$ such that
$(A_0 \wedge B_0)^+ \models_{{\mathcal T}_0} c_1 R_1 d_1$. 
\end{itemize}

\medskip
\noindent {\bf Case 1:} $(A_0 \wedge B_0)^+ \models a R b$. Then we
are done. 

\noindent {\bf Case 2:} $A_0 \wedge B_0$ is unsatisfiable. In this case 
(i) and (ii) hold for $T = \emptyset$.

\medskip
\noindent {\bf Case 3:} $A_0 \wedge B_0$ is satisfiable, $(A_0 \wedge
B_0)^+ \not\models a R b$, and there exists $(c_1 R_1 d_1) \in {\mathcal D}$ such that
$(A_0 \wedge B_0)^+ \models_{{\mathcal T}_0} c_1 R_1 d_1$. 
Then $A_0 \wedge B_0$ is logically equivalent in ${\mathcal T}_0$ with 
$A_0 \wedge B_0 \wedge c_i R_i d_i$. 
If it is not the case that by adding $c_i R_i d_i$ all premises of some 
rule in ${\mathcal H}$ become true we repeat the procedure for 
${\mathcal D}_1 = {\mathcal D} \backslash (c_1 R_1 d_1)$:  
Again in this case $A_0 \wedge B_0 \wedge
\bigwedge_{(c R' d) \in {\mathcal D}_1} \neg (c R' d) \wedge \neg (a R
b)  \models_{{\mathcal T}_0} \perp$ (if it has a model then 
$A_0 \wedge B_0 \wedge {\mathcal H} \wedge \neg (a R b)$ has one), 
and as before, using convexity we infer that 
either $A_0 \wedge B_0 \models a R b$ (which cannot be the case) or 
$A_0 \wedge B_0$ is unsatisfiable (which cannot be the case) 
or there exists $c_2 R_2 d_2 \in {\mathcal D}_1$ with 
$A_0 \wedge B_0 \models_{{\mathcal T}_0} c_2 R_2 d_2$. 
We can repeat the process until all the premises of some clause 
in ${\mathcal H}$ are proved to be entailed by $A_0 \wedge B_0$. 
Let $C = \bigwedge_{i = 1}^n  c_i R_i d_i \rightarrow c R d$ be such a clause.

\smallskip
\noindent 
{\bf Case 3a.} Assume that $C$ contains only constants occurring in 
$A$ or only constants occurring in $B$.  
Then $A_0 \wedge B_0 \wedge {\mathcal H}$ is equivalent with respect to 
${\mathcal T}_0$ with 
$A_0 \wedge B_0 \wedge ({\mathcal H} \backslash C) \wedge c R d$. 
By the induction hypothesis for  
$A_0' \wedge B_0' = A_0 \wedge B_0 \wedge c R d$ and 
${\mathcal H'} = {\mathcal H} \backslash \{ C \}$,  we know that 
there exists $T'$ such that 
$A'_0 \wedge B'_0 \wedge ({\mathcal H'} \backslash {\mathcal H'}_{\sf mix}) 
\wedge {\mathcal H'}_{\sf sep} \wedge \neg (a R b) \models \perp$, and (ii) holds too.

\smallskip
\noindent Then, for $T = T'$, 
$A'_0 \wedge B'_0 \wedge ({\mathcal H'} \backslash {\mathcal H'}_{\sf mix}) 
\wedge {\mathcal H'}_{\sf sep}$ is logically equivalent to 
$A_0 \wedge B_0 \wedge ({\mathcal H} \backslash {\mathcal H}_{\sf mix}) \wedge 
{\mathcal H}_{\sf sep}$, so (i) holds. 

\smallskip  
\noindent In order to prove (ii), note that, by definition, 
${\mathcal H}'_{\sf mix} = {\mathcal H}_{\sf mix}$ and  
${\mathcal H}'_{\sf sep} = {\mathcal H}_{\sf sep}$. 
By the induction hypothesis, 
$A'_0 \wedge B'_0 
\wedge ({\mathcal H'} \backslash {\mathcal H'}_{\sf mix}) \cup 
{\mathcal H'}_{\sf sep}$ is logically equivalent to a corresponding 
conjunction ${\overline A'_0} \wedge {\overline B'_0}$ containing 
as conjuncts all literals in $A'_0$ and $B'_0$ and all conclusions of 
rules in ${\mathcal H'} \backslash {\mathcal H'}_{\sf mix}$ and 
${\mathcal H'}_{\sf sep}$. 
On the other hand, $A'_0 \wedge B'_0$ is logically equivalent to 
$A_0 \wedge B_0 \wedge (c R d)$, where $(c R d)$ is the 
conclusion of the rule $C \in {\mathcal H} \backslash {\mathcal H}_{\sf mix}$. 
This proves (ii). 

\medskip
\noindent 
{\bf Case 3b.} Assume now that $C$ is mixed, for instance that 
$c_1, \dots, c_n, c$ are 
constants in $A$ and $d_1, \dots, d_n, d$ are constants in $B$. 
Assume that $C$ is obtained from an instance of a clause of the form 
$\bigwedge_{i = 1}^n x_i R_i s_i({\overline y}) \rightarrow 
f(x_1, \dots, x_n) R g({\overline y}).$ (The case when $C$ corresponds to 
an instance of a monotonicity axiom is similar.) 
This means that there exist $c \approx f(c_1, \dots, c_n) \in D_A$ 
and $d_i \approx s_i({\overline e}), d \approx g({\overline e}) \in
D_B$. 
In particular, $f$ is a $A$-function symbol and $g$ a $B$-function
symbol; since $f$ and $g$ occur together in an axiom in $\K$ they are 
considered to be shared. 
The clause $C$ was chosen such that for each premise $c_i R_i d_i$ of $C$, 
$A_0 \wedge B_0 \models_{{\mathcal T}_0} c_i R_i d_i$, and  ${\mathcal T}_0$ is $P$-interpolating.  
Thus, there exist terms $t_1, \dots, t_n$ containing only constants 
common to $A_0$ and $B_0$ such that for all $i \in \{ 1, \dots, n \}$
\begin{eqnarray}
A_0 \wedge B_0 \models_{{\mathcal T}_0} c_i R_i t_i \wedge t_i R_i d_i. \label{interp-equation} 
\end{eqnarray}
Let $c_{f(t_1, \dots, t_n)}$ be a new constant, denoting the term 
$f(t_1, \dots, t_n)$, and let 
$$C_A  =  \bigwedge_{i = 1}^n c_i R_i t_i  {\rightarrow} 
c R c_{f(t_1, \dots, t_n)} \quad \text{ and } \quad 
C_B  =  \bigwedge_{i = 1}^n t_i R_i d_i {\rightarrow} 
c_{f(t_1, \dots, t_n)} R d.$$ 
Thus, $C_A$ corresponds to the monotonicity axiom 
$$\displaystyle{\bigwedge_{i = 1}^n c_i R_i t_i 
{\rightarrow} f(c_1, \dots, c_n) R f(t_1, \dots, t_n)},$$ whereas  
$C_B$ corresponds to the rule 
$$\displaystyle{\bigwedge_{i = 1}^n t_i R_i 
s_i({\overline e}) {\rightarrow} f(t_1, \dots, t_n) R g({\overline e})}.$$ 
As $R$ is transitive, by~(\ref{interp-equation}) the following holds:

\noindent $\begin{array}{@{}lll}
A_0 \wedge B_0 \wedge C_A \wedge C_B \wedge \neg a R b &
\models\!\!\!|_{{\mathcal T}_0} & A_0 \wedge B_0 \wedge (\displaystyle{\bigwedge_{i
  = 1}^n c_i R_i t_i }\wedge C_A) \wedge 
(\displaystyle{\bigwedge_{i = 1}^n  t_i R_i d_i} \wedge C_B) \\
& & \wedge \neg a R b\\
& \models\!\!\!|_{{\mathcal T}_0} &   A_0 \wedge B_0 \wedge  c R
c_{f(t_1, \dots, t_n)} \wedge c_{f(t_1, \dots, t_n)} R d \wedge \neg a
R b \\
& \models_{{\mathcal T}_0} &   A_0 \wedge B_0 \wedge  c R d   \wedge
\neg a R b
\end{array}$

\smallskip
~~~~~(where $\models\!\!\!|_{{\mathcal T}_0}$ stands for logical equivalence with respect to 
${\mathcal T}_0$). 

\medskip
\noindent Hence,
$A_0 \wedge B_0 \wedge C_A \wedge C_B \wedge ({\mathcal H} \backslash
C) \wedge \neg a R b 
 \models_{{\mathcal T}_0}    A_0 \wedge B_0 \wedge  c R d \wedge
 ({\mathcal H} \backslash C) \wedge \neg a R b.$ 

\smallskip
\noindent On the other hand, 
as $A_0 \wedge B_0 \models_{{\mathcal T}_0} \bigwedge_{i = 1}^n c_i R_i d_i$,
$A_0 \wedge B_0 \wedge {\mathcal H}$ is logically equivalent with 
$ A_0 \wedge B_0 \wedge c R d \wedge ({\mathcal H} \backslash C)$, so 
$A_0 \wedge B_0 \wedge C_A \wedge C_B \wedge ({\mathcal H} \backslash
C) \wedge \neg a R b \models_{{\mathcal T}_0} \perp$. 

\smallskip
\noindent 
By the induction hypothesis for 
$A_0 \wedge B_0 \wedge c R c_{f(t_1, \dots, t_n)} \wedge c_{f(t_1, \dots, t_n)} R d$ and ${\mathcal H}' = {\mathcal H} \backslash C$ we know that there exists a set 
$T'$ of terms such that 
$$A_0 \wedge B_0 \wedge c R c_{f(t_1, \dots, t_n)} \wedge c_{f(t_1,
  \dots, t_n)} R d \wedge ({\mathcal H}' \backslash {\mathcal H'}_{\sf
  mix}) \wedge {\mathcal H}'_{\sf sep} \wedge \neg a R b 
\models \perp,$$ and also (ii) holds. 
Then (i) holds for $T = T' {\cup} \{ t_1, \dots, t_n \}$. 
(ii) can be proved similarly using the induction hypothesis.
\QED

\begin{thm}
Assume that ${\cal T}_0$ is convex and $P$-interpolating with
respect to  $P \subseteq {\sf Pred}$, and that the ground
satisfiability w.r.t.\ $\T_0$ is decidable and the interpolating terms can
be effectively computed. 
Assume that ${\cal T} = {\cal T}_0 \cup {\cal K}$ 
  is a $\Psi$-local extension of ${\cal T}_0$ with a set of clauses ${\cal K}$ 
which only contains combinations of clauses of type~(\ref{general-form}).
Then $\T$ is also $P$-interpolating and the interpolating terms can be
effectively computed.  
\label{thm-pint}
\end{thm}
{\em Proof:} 
We prove that if $A$, $B$ are conjunctions of literals 
and $A({\overline c}, \overline{a}_1, a) \wedge 
B({\overline c}, \overline{b}_1, b) \models_{\T} a R b$ 
where $R \in P$, then there exists a term $t$ containing only the
constants common to $A$ and $B$ and only function symbols which are
shared by $A$ and $B$, such that $A({\overline c}, \overline{a}_1, a)
\wedge B({\overline c}, \overline{b}_1, b) \models_{\T} a R t ~\wedge~ t R b$. 
We can restrict w.l.o.g.\ to a purified and flattened conjunction 
of unit clauses $A_0 \wedge
B_0 \wedge {\sf Def} \wedge \neg (a R b).$ With the notation used
on page~\pageref{page}, by Theorem~\ref{lemma-rel-transl} we have: 

\smallskip
\noindent $A_0 \wedge B_0 \wedge {\sf Def} \wedge \neg (a R b) 
{\models}_{{\cal T}} {\bot} \text{ iff } 
{\cal K}_0 \wedge A_0 \wedge B_0 \wedge {\sf Con}[D_A \wedge D_B]_0 
\wedge \neg (a R b) {\models}_{{\cal T}_0} 
{\bot}.$

\smallskip
\noindent By Proposition~\ref{sep} (ii), there exists a set $T$ of $(\Sigma_0 \cup C)$-terms containing 
only constants common to $A_0$ and $B_0$ such that 
${\cal H} = \K_0 \wedge {\sf Con}[D_A \wedge D_B]_0$
can be separated as described in Proposition~\ref{sep}, 
$A_0 \wedge B_0 \wedge ({\cal H} \backslash {\cal H}_{\sf mix}) 
\wedge {\cal H}_{\sf sep} \wedge \neg a R b$ is logically equivalent w.r.t.\ ${\cal T}_0$ with 
a separated  conjunction of ground literals 
${\overline A}_0 \wedge {\overline B}_0 \wedge \neg a R b$, which is
therefore unsatisfiable, so ${\overline A}_0
\wedge {\overline B}_0 \models a R b$.
From the $P$-interpolation property in $\T_0$, there exists a term
containing the shared constants such that ${\overline A}_0 \wedge
{\overline B}_0 \models_{\T_0} a R t ~\wedge~ t R b$. 
If we now replace all constants $c_{f(t_1, \dots, t_n)}$ introduced in
the purification process or in the separation process with the terms
they denote, we obtain $A \wedge B \models_{\cal T} a R t \wedge
t R b$. Since all intermediate terms $t_i$ contain only shared symbols
and the function symbol $f$ is shared by $A$ and $B$, all terms 
$f(t_1, \dots, t_n)$, hence also $t$ contain only symbols shared by
$A$ and $B$. \QED

\

 \noindent 
We obtain the following procedure for $P$-interpolation if $A \wedge B
\models_{\T} a R b$: 

\begin{description}
\item[Step 1: Preprocess] Using locality, flattening and purification we obtain a 
set ${\cal H} \wedge A_0 \wedge B_0$ of formulae in the base theory, where 
${\cal H}$ is as in Proposition~\ref{sep}. 

\item[Step 2:]  $\Delta := {\sf T}$. {\bf Repeat as long as $A_0 \wedge B_0 \wedge \Delta \not\models a R
b$:}
  \\
Let $C {\in} {\cal H}$ whose premise is entailed by 
$A_0 {\wedge} B_0 {\wedge} \Delta$. \\
If $C$ is not mixed, 
move $C$ to ${\cal H}_{\sf sep}$ and add its conclusion
to $\Delta$.\\
If $C$ is mixed, compute terms $t_i$ which separate the premises in $C$, 
and separate the clause into an instance $C_1$ of monotonicity and 
an instance $C_2$ of a clause in ${\cal K}$ as in the proof of
Proposition~\ref{sep}. 
Remove $C$ from ${\cal H}$, and add $C_1, C_2$ to ${\cal H}_{\sf sep}$ 
and their conclusions  to $\Delta$. 

\item[Step 3: Compute separating term.] Compute a separating term for 
$A_0 \wedge B_0 \wedge \Delta \models a R b$ in ${\cal T}_0$, 
and construct an interpolant for the extension as explained in the
proof of Theorem~\ref{thm-pint}. 
\end{description}

\section{Example: Semilattices with monotone operators}
\label{semilattices}

We will now analyze $\leq$-interpolation properties for theories of
semilattices with monotone operators.

\subsection{The theory ${\sf SLat}$ of semilattices}

We define the theory ${\sf SLat}$ of semilattices and show that conditions ${\bf A1}$ and ${\bf A2}$ are satisfied for 
${\sf SLat}$. 

\smallskip
\noindent A semilattice $(S, \swedge)$ is set $S$ with a binary
operation $\swedge$ which is associative, commutative and
idempotent. One can equivalently regard semilattices as partially
ordered sets $(S, \leq)$, in which infima $a_1 \swedge \dots \swedge
a_n$ of finite non-empty subsets $\{ a_1, \dots, a_n \} \subseteq S$ exist;
then $a \leq b$ iff $a \swedge b = a$. 

\smallskip
\noindent The theory ${\sf SLat}$  of semilattices can be axiomatized
by equations (associativity, commutativity and idempotence of
$\swedge$), therefore it clearly is $\approx$-convex: Convexity
w.r.t.\ $\leq$ follows from the fact that $x \leq y$ iff $(x
\swedge y) \approx x$.

\begin{lemma}
Ground satisfiability w.r.t.\ ${\sf SLat}$ is decidable.
\end{lemma}
{\em Proof:} 
This is a consequence of the fact that the theory of semilattices
${\sf SLat}$ is a local extension of the theory of pure equality 
-- this follows from a result on the locality of lattices by
Skolem \cite{skolem}, or by results in \cite{Ganzinger-01-lics}, since
every partial semilattice weakly embeds into a total one.
\footnote{There are also other justifications for the decidability of ground
satisfiability w.r.t.\ ${\sf SLat}$, leading to different types of
decision procedures: 
${\sf SLat} = ISP(S_2)$, where $S_2$ is the 2-element
 semilattice, i.e.\ 
 every semilattice is isomorphic to a
 sublattice of a power of $S_2$ -- or, alternatively, that every semilattice is isomorphic
 to a semilattice of sets. Thus, checking satisfiability w.r.t.\ ${\sf SLat}$
 can be reduced to checking satisfiability w.r.t.\ $S_2$ and
 ultimately to propositional reasoning.} \QED

\smallskip
\noindent The theory ${\sf SLat}$ is 
$\leq$-interpolating, therefore also $\approx$-interpolating
(cf.\ \cite{peuter-sofronie}; we present the proof since 
it indicates how the intermediate terms can be computed):

\begin{lem}
  The theory ${\sf SLat}$ of semilattices is 
  $\leq$-interpolating.
\label{slat-p-interp}
\end{lem}

\noindent
{\em Proof:} This is a constructive proof based on the fact that 
${\sf SLat} = ISP(S_2)$, where $S_2$ is the 2-element
 semilattice, i.e.\ 
 every semilattice is isomorphic to a
 sublattice of a power of $S_2$ -- or, alternatively, that every semilattice is isomorphic
 to a semilattice of sets. 
We prove that the theory of semilattices is $\leq$-interpolating, 
i.e. that if $A$ and $B$ are two conjunctions of literals 
and $A \wedge B \models_{\sf SLat} a \leq b$, where $a$ is a 
constant occurring in $A$ 
 and $b$ a constant occurring in $B$, 
then 
there exists a term containing only common constants in $A$ and $B$ such that 
$A \models_{\sf SLat} a \leq t$ and $A \wedge B \models_{\sf SLat} t \leq b$. 
We can assume without loss of generality that $A$ and $B$ consist only of 
atoms:   Indeed, assume that 
$A \wedge B = A_1 \wedge \dots \wedge A_n \wedge \neg A'_1 \wedge
\dots \wedge \neg A'_m$, where $A_1, \dots, A_n, A'_1, \dots, A'_m$ are
atoms. 
Then the following are equivalent: 
\begin{itemize}
\item $A \wedge B \models_{\sf SLat} a \leq b$ 
\item $\models_{\sf SLat} A \wedge B \rightarrow a \leq b $
\item $\models_{\sf SLat} \neg A_1 \vee \dots \vee \neg A_n \vee A'_1
  \vee \dots \vee A'_m \vee a \leq b$
\item $\models_{\sf SLat} (A_1 \wedge \dots \wedge A_n) \rightarrow A'_1
  \vee \dots \vee A'_m \vee a \leq b$
\item $A_1 \wedge \dots \wedge A_n \models_{\sf SLat} A'_1
  \vee \dots \vee A'_m \vee a \leq b$
\end{itemize}
Since the theory of semilattices is convex w.r.t. $\leq$ and
$\approx$, it follows that if $A \wedge B \models_{\sf SLat} a \leq b$
then either 
\begin{itemize}
\item[(a)] $A_1 \wedge \dots \wedge A_n \models_{\sf SLat} A'_j$ for
some $j \in \{ 1, \dots, m \}$ or 
\item[(b)] $A_1 \wedge \dots \wedge A_n
\models_{\sf SLat} a \leq b$. 
\end{itemize}
It is easy to see that in case (a), $A \wedge B \models \perp$. Then
the conclusion follows immediately. 
We therefore consider the case when $A$ and $B$ consist
only of atoms. 

As ${\sf SLat} = ISP(S_2)$, in ${\sf SLat}$ the same Horn sentences
are true as in the 2-element semilattice $S_2$. Thus, 
$A \wedge B \models_{\sf SLat} a \leq b$ iff $A \wedge B \models_{S_2}
a \leq b$, so we can reduce such a test to entailment in propositional
logic. 

It follows that 
$A \wedge B \models_{\sf SLat} a \leq b$ if and only if the following conjunction of literals in 
propositional logic is unsatisfiable: 
$$\begin{array}{@{}c@{}c@{}c} 
N_A{:} \left\{ \begin{array}{@{}r@{}c@{}l}
P_{e_1 \swedge e_2} & \leftrightarrow & P_{e_1} \wedge P_{e_2} \\
P_{e_1} & \leftrightarrow & P_{e_2} ~~~ e_1 \approx e_2
\in A\\
P_{e_1} & \rightarrow & P_{e_2} ~~~ e_1 \leq e_2
\in A \\
\text{ for all } e_1, & e_2 & \text{ subterms in } A
\end{array} \right. & ~ & 
N_B{:} \left\{ \begin{array}{@{}rcl}
P_{g_1 \swedge g_2} & \leftrightarrow & P_{g_1} \wedge P_{g_2} \\
 P_{g_1} & \leftrightarrow & P_{g_2} ~~~ g_1 \approx g_2 \in B\\
 P_{g_1} & \rightarrow & P_{g_2} ~~~ g_1 \leq g_2 \in B\\
\text{ for all } g_1, & g_2 & \text{ subterms in } B 
\end{array} \right.\\
P_a & \hspace{0.5cm} & \neg P_b \\
\end{array}$$

\noindent 
i.e.\ $A \wedge B \models_{\sf SLat} a \leq b$ if and only if 
$(N_A \wedge P_a) \wedge (N_B \wedge \neg P_b) \models \perp$, 
where 
$N_A$ and $N_B$ are sets of Horn clauses in which each clause contains
a positive literal. 

\noindent We show that if $A \land B \models_{\sf SLat} a \leq b$ holds, then for
the term 
\[ t := \bigswedge\{e \mid A \models_{\sf SLat} a \leq e, e \text{ common subterm
  of } A \text{ and } B\}\]
the following hold: 
\begin{itemize}
\item[(i)] $A \models_{\sf SLat} a \leq t$, and 
\item[(ii)] $A \land B \models_{\sf SLat} t \leq b$. 
\end{itemize}
This means that for the theory of semilattices we have a property
stronger than $\leq$-interpolability. 

\noindent 
Every $e \in T = \{e \mid A \models_{\sf SLat} a \leq e, e \text{ common subterm
  of } A \text{ and } B\}$ corresponds to the positive unit clause
$P_e$ (where $P_e$ is a propositional variable common to
$N_A$ and $N_B$) which can be derived from $N_A$ using 
ordered resolution (with the ordering described above).  

\smallskip
\noindent 
It is clearly the case that $A \models_{\sf SLat} a \leq t$, because 
$N_A \wedge P_a \wedge \neg P_{t} \wedge (P_t \leftrightarrow
\bigwedge_{e \in T} P_e)$ is unsatisfiable. 
Thus, (i) holds.

\smallskip
\noindent 
For proving (ii), observe that by saturating $N_A \wedge P_a$ under
ordered resolution we obtain the following
kinds of clauses which can possibly lead to $\perp$ after inferences
with $N_B \wedge \neg P_b$ 
(and thus to the consequence $a \leq b$ together with
$B$):
\begin{itemize}
\item[(a)] $P_{e_k}$ positive unit clauses s.t.\ $e_k$ contains 
symbols common to $A$ and $B$, for $k \in \{ 1, \dots, l \}$.
\item[(b)] $\bigwedge_{j = 1}^{n_i} P_{c_{ij}} \rightarrow P_{d_i}$, where $c_{ij}$ and $d_i$ 
are common symbols, such that for all $i$, $j$ and $k$ we have 
$ c_{ij} \neq e_k$ and $d_i \neq e_k$, for $i \in \{ 1, \dots, p \}$.
\end{itemize} 
Other types of clauses may appear too, but they can not be used to obtain $a \leq b$:

To see that clauses where some $c_{ij} = e_k$ are not necessary to
derive the consequence $a \leq b$, note that if $P_{e_k}$ is a
positive unit literal and we have the clause $(P_{e_k} \land \bigwedge
P_{c_{ij}}) \rightarrow P_{d_i}$, then by resolution we get as an
inference $\bigwedge P_{c_{ij}} \rightarrow P_{d_i}$. It is easy to see
that 
$(P_{e_k} \land \bigwedge P_{c_{ij}}) \rightarrow
P_{d_i}$ is redundant
in the presence of $\bigwedge P_{c_{ij}} \rightarrow P_{d_i}$. 
In the same way, 
clauses of the form $\bigwedge P_{c_{ij}} \rightarrow P_{e_k}$ (i.e. clauses of type (b) where  $d_i = e_k$) are redundant in the presence of 
$P_{e_1}, \dots, P_{e_l}$.
For the proof of (ii) one needs to consider separately the case in
which none of the $P_{d_i}$ is needed to derive $\perp$ together with
$N_B$ (and thus the consequence $a \leq b$) 
and the case when some $P_{d_i}$ are needed. 

\medskip

\noindent
{\bf Case 1:} None of the $P_{d_i}$ is needed to derive $\perp$ together with
$N_B$ (and thus the consequence $a \leq b$). 
We know that $N_A \models P_a \rightarrow \bigwedge_{k = 1}^l P_{e_k}$.
From this it follows that $A \models a \leq \bigswedge_{k=1}^l e_k$. 

\noindent For $A \land B \models a \leq b$ to be true, 
$\bigwedge_{k = 1}^l P_{e_k} \wedge N_B \wedge \neg P_b$ must be
unsatisfiable, so there has to be a subset 
$S \subseteq \{1,...,l\}$ such that $\bigwedge_{k \in S} P_{e_k}
\wedge N_B \wedge \neg P_b$. 
This means that $B \models \bigswedge_{s \in S} e_s \leq b$.  
But then, since $\bigswedge_{k=1}^l e_k \leq \bigswedge_{s \in S} e_s$, 
it follows that $B \models \bigswedge_{k=1}^l e_k \leq b$, and therefore also $A \land B \models \bigswedge_{k=1}^l e_k \leq b$.

\medskip

\noindent
{\bf Case 2:} Some $P_{d_i}$ are needed to derive $\perp$ from $N_B
\wedge \neg P_b$. 
Again, we know that $N_A \models P_a \rightarrow \bigwedge_{k = 1}^l P_{e_k}$
(hence $A \models a \leq \bigswedge_{k=1}^l e_k$). 

\noindent 
For $A \land B \models a\leq b$ to be true, i.e.\ 
$(N_A \wedge P_a) \wedge (N_B \wedge \neg P_b)$ to be unsatisfiable, 
there have to be subsets $S_1 \subseteq \{1, \dots,l\}$ and 
$S_2 \subseteq \{1, \dots,p\}$ such that $N_B \wedge \bigwedge_{k \in
  S_1} P_{e_k} \wedge \bigwedge_{i \in
  S_2} ((\bigwedge_j P_{c_{ij}}) \rightarrow P_{d_i}) \wedge \neg P_b$ is unsatisfiable. 
 
\smallskip
\noindent Let $N_{AB} := N_B \wedge \bigwedge_{k \in
  S_1} P_{e_k} \wedge \bigwedge_{i \in
  S_2} ((\bigwedge_j P_{c_{ij}}) \rightarrow P_{d_i})$. 

\smallskip
\noindent 
We know that $N_{B} \wedge \bigwedge_{k \in
  S_1} P_{e_k} \wedge \neg P_b$ is satisfiable. 
Assume that there is no $c_{ij}$ such that $N_B \wedge \bigwedge_{k \in
  S_1} P_{e_k} \wedge \neg P_b \models P_{c_{ij}}$. 
Then for every $c_{ij}$, $N_B \wedge \bigwedge_{k \in
  S_1} P_{e_k} \wedge \neg P_b \wedge \neg P_{c_{ij}}$ is
satisfiable.  Since all clauses in $N_b \wedge \bigwedge_{k \in
  S_1} P_{e_k} \wedge \neg P_b$ are Horn clauses, it follows that 
$N_B \wedge \bigwedge_{k \in
  S_1} P_{e_k} \wedge \neg P_b \wedge \bigwedge_{i,j} \neg P_{c_{ij}}$ is
satisfiable.  Every model of $N_B \wedge \bigwedge_{k \in
  S_1} P_{e_k} \wedge \neg P_b \wedge \bigwedge_{i,j} \neg P_{c_{ij}}$
is a model of $N_{AB} \wedge \neg P_b$. It would therefore follow 
that $N_{AB} \wedge \neg P_b$ is satisfiable, which is a
contradiction.

\smallskip
\noindent  Thus, there exists at least one $c_{ij}$ such that $N_B \wedge \bigwedge_{k \in
  S_1} P_{e_k} \wedge \neg P_b \models P_{c_{ij}}$. 

\smallskip
\noindent 
We can add
$P_{c_{ij}}$ to this set of clauses and repeat the reasoning for the
set of clauses obtained this way as long as we still have one clause 
of the form $((\bigwedge_j P_{c_{ij}}) \rightarrow P_{d_i})$ in
$N_{AB}$ such that there exists at least one $c_{ij}$ such that
$P_{c_{ij}}$ was not added to $N_{AB}$.

\smallskip
\noindent 
Then there has to be a sequence $(d_{i_1j})_{j \in J_1}, (d_{i_2j})_{j
  \in J_2}, ... , (d_{i_nj})_{j \in J_n}$ such that:
\begin{itemize}
\item $P_{d_{i_1j}}$ can be derived from $N_{AB} {\wedge} \bigwedge
    P_{e_k}$, for all $j \in J_1$, 
\item $P_{d_{i_2j}}$ can be derived from $N_{AB} {\wedge} \bigwedge
    P_{e_k} {\wedge} \bigwedge_{k \in J_1} P_{d_{i_1k}}$, for all $j \in J_2$, 
\item $P_{d_{i_3j}}$ can be derived from $N_{AB} {\wedge} \bigwedge
    P_{e_k} {\wedge} \bigwedge_{k \in J_1} P_{d_{i_1k}} {\wedge} \bigwedge_{k
      \in J_2} P_{d_{i_2k}} $,  for all $j \in J_3$, \\
$\dots$
\item $P_{d_{i_nj}}$ can be derived from $N_{AB} {\wedge} \bigwedge_{k}
    P_{e_k} {\wedge} \bigwedge_{k \in J_1} P_{d_{i_1k}} {\wedge} \dots 
    \bigwedge_{k \in J_{n-1}} P_{d_{i_{n-1}k}} $,  for all $j \in J_n$, 
\item $P_b$ can be derived from $N_{AB} {\wedge} \bigwedge
    P_{e_k} {\wedge} \bigwedge_{k \in J_1} P_{d_{i_1k}} {\wedge} \dots {\wedge}
    \bigwedge_{k \in J_n} P_{d_{i_nk}} $.  
\end{itemize}
But then 
$A \wedge B \models \bigswedge e_k \leq d_{i_1l}$, for all $l \in J_1$, 
hence 
$A \wedge B \models \bigswedge e_k \leq \bigswedge_{l \in J_1} d_{i_1l}$, hence 
$A \wedge B \models (\bigswedge e_k \swedge \bigswedge_{l \in J_1} d_{i_1l}) \approx \bigswedge e_k$.

\smallskip
\noindent  
Therefore, as  
$A \wedge B \models (\bigswedge e_k \swedge \bigswedge_{l \in J_1}
d_{i_1l}) \leq d_{i_2j}$, for all $j \in J_2$,  we conclude that 
$A \wedge B \models \bigswedge e_k \leq \bigswedge_j d_{i_2j}$.

\smallskip
\noindent 
Similarly it can be proved that 
$A \wedge B \models \bigswedge e_k \leq \bigswedge_{j \in J_n} d_{i_nj}$, 
and finally that $A \wedge B \models \bigswedge e_k \leq b$. \qed

\medskip
\noindent
We illustrate the computation of intermediate terms  on an example.
\begin{example}
Let $A=\{ a_1 \leq c_1,~ c_2 \leq a_2,~ a_2 \leq c_3\}$ and 
$B=\{c_1 \leq b_1,~ b_1 \leq c_2,~ c_3 \leq b_2\}$. 
It is easy to see that $A \wedge B \models a_1 \leq b_2$. 
We can find an intermediate term by using the methods described in the
proof of Lemma~\ref{slat-p-interp}: 
We saturate the set of clauses 

\smallskip
$N_A \wedge P_{a_1} = (P_{a_1} \rightarrow P_{c_1}) \wedge
(P_{c_2} \rightarrow P_{a_2})  \wedge (P_{a_2} \rightarrow P_{c_3})
\wedge P_{a_1}$

\smallskip
\noindent under ordered resolution, in which the propositional variables $P_{a_1}, P_{a_2}$ are
larger than $P_{c_1}, P_{c_2}, P_{c_3}$. 
This yields the clauses $P_{c_1}$ and $P_{c_2} \rightarrow P_{c_3}$
containing shared propositional variables. $(N_A \wedge P_{a_1}) \wedge
(N_B \wedge \neg P_{b_2})$ is unsatisfiable iff $N_B \wedge \neg
P_{b_2} \wedge P_{c_1} \wedge (P_{c_2} \rightarrow P_{c_3})$ is
unsatisfiable. 
Indeed $t = c_1$ is an intermediate term, as $A \models a_1 \leq c_1$
and $A \wedge B \models c_1 \leq b_2$. 
Note that $N_B \wedge \neg P_{b_2} \wedge P_{c_1}$ is satisfiable, so 
$B \not\models c_1 \leq b_2$. 
Moreover, we only need $P_{c_2} \rightarrow P_{c_3}$ in addition
to $N_B \cup \neg P_{b_2}$ to derive $\perp$, thus $A \wedge B \models
c_1 \leq b_2$ and the clause $P_{c_2} \rightarrow P_{c_3}$ obtained
from $N_A$ is really needed for this. \QED
\end{example}
\subsection{Semilattices with operators} 

We define the theory of semilattices with
monotone operators in a set $\Sigma$ possibly satisfying additional
properties and 
show that conditions ${\bf A3}$ and ${\bf A4}$ are satisfied for 
this theory. 

\smallskip
\noindent Let $\Sigma$ be a set of unary\footnote{We assume that the
  function symbols are unary to simplify the presentation, and because
  in the applications to description logics we need only unary
  function symbols. All the
  results can be extended to function symbols of higher arity.} function symbols. We consider 
the extension ${\sf SLat}_{\Sigma} = {\sf SLat} \cup {\sf
  Mon}(\Sigma)$ of ${\sf SLat}$ with new function
symbols in $\Sigma$ satisfying the monotonicity axioms ${\sf Mon}_{\Sigma} = \bigcup_{f \in \Sigma} {\sf Mon}(f)$,
where: 

\smallskip
${\sf Mon}(f) \quad \quad \forall x, y (x \leq y \rightarrow f(x) \leq
f(y))$

\smallskip
\noindent and also extensions  ${\sf SLat} \cup {\sf
  Mon}(\Sigma) \cup \K$, where $\K$ is a set of axioms of the form: 
\begin{align}
\forall x ~~~  & \phantom{y \leq g(x) \rightarrow}  f(x) \leq g(x)  \label{inclusion} \\
\forall x, y ~~~ &  y \leq g(x) \rightarrow  f(y) \leq
h(x) \label{composition} 
\end{align}

\smallskip
\noindent where $f, g, h \in \Sigma$, not necessarily all different.


\medskip
\noindent Clearly, condition ${\bf A3}$ is satisfied for the theory of
semilattices with operators defined above. We show that condition
${\bf A4}$ holds as well.
\begin{lem}
The following extensions satisfy a locality property: 
\begin{enumerate}[(i)]
\item The theory of semilattices ${\sf SLat}$ is local. 
\item ${\sf SLat} \cup {\sf Mon}_{\Sigma}$ is a local extension
  of ${\sf SLat}$. 
\item ${\sf SLat} \cup {\sf Mon}_{\Sigma} \cup \K$ is a $\Psi$-local extension
  of ${\sf SLat}$, where $\Psi$ is the closure operator on ground terms
  defined as follows: 
\begin{align*}
\Psi(G) =&~ \bigcup_{i \geq 0} \Psi^i(G), \text{ with~ } 
\Psi^0(G) =\textsf{est}(G) \text{ (the set of ground terms in $G$}\\[-3ex]
& ~~~~~~~~~~~~~~~~~~~~~~~~~~~~~~~ \text{starting with extension functions), and} & \\
\Psi^{i+1}(G) =&~ \{ h(c) \mid \forall x (g(x) \leq h(x)) \in \K \text{ and } g(c) \in \Psi^i(G)\} \cup \\
	&~ \{ g(c) \mid \forall x (g(x) \leq h(x)) \in \K \text{ and } h(c) \in \Psi^i(G)\}\cup \\
	&~ \{ h(c) \mid \forall x, y  (y \leq g(x) \rightarrow f(y) \leq
        h(x)) \in \K \text{ and } g(c) \in \Psi^i(G)\}\cup \\
	&~ \{ g(c) \mid \forall x, y (y \leq g(x) \rightarrow f(y) \leq h(x)) \in \K \text{ and } h(c) \in \Psi^i(G)\}.
\end{align*}
\end{enumerate}
\label{lemma-psi}
\end{lem}
{\em Proof:} (i) follows from a result on the locality of lattices by
Skolem \cite{skolem}, or by results in \cite{Ganzinger-01-lics}, since
every partial semilattice weakly embeds into a total one.\\
(ii) follows from results in \cite{Ihlemann-Sofronie-ismvl,sofronie-ihlemann-ismvl-07}.
(iii) Since the axioms in $\K$ are not always linear, we use
the locality criterion for non-linear sets of clauses mentioned in Theorem~\ref{rel-loc-embedding}, 
and the fact that every semilattice $P = (S, \swedge, \{ f \}_{f \in \Sigma})$ with
partially defined monotone operators satisfying the axioms
$\K$, and with the property that if a variable occurs in two terms $g(x),
h(x)$ in a clause in $\K$, then for every $s \in S$,  $g(s)$ is
defined iff $h(s)$ is defined, weakly embeds into a semilattice with
totally defined operators satisfying $\K$, which was proved in 
Lemma~4.5 from \cite{sofronie-fuin-2017}. \QED


\medskip
\noindent Given two sets of conjunctions of ground literals $A$ and
$B$ over the signature of semilattices with operators, we consider the
lattice operation $\swedge$ to be interpreted and the function
symbols in $\Sigma$ to be uninterpreted. Let $\Sigma_A$ be the function symbols in
  $\Sigma$ occurring in $A$ and $\Sigma_B$ those occurring in $B$.
 We consider the following variants for ``shared uninterpreted function symbols'': 
\begin{itemize}
\item {\em Intersection-sharing:} The shared function symbols of $A$
  and $B$ are the function symbols in $\Sigma_A \cap \Sigma_B$. 
\item {\em $\Theta_\K$-sharing} (as defined in
 Example~\ref{ex-shared}):  
$\Theta_\K(\Sigma_A) = \bigcup_{f \in \Sigma_A} \{ g \in \Sigma \mid f
\sim^*_{\K} g \}$, where $\sim^*_{\K}$ is the equivalence relation
induced by $\sim_{\K}$ (with $f \sim_{\K} g$ iff $f, g$ occur in the
same clause in $\K$); $\Theta_\K(\Sigma_B)$ is
defined analogously. The $\Theta_\K$-shared function symbols are the function 
symbols in $\Theta_{\K}(\Sigma_A) \cap \Theta_\K(\Sigma_B)$. 
\end{itemize} 
\begin{thm}
For every set $\K$ containing clauses of the form~(\ref{inclusion})
and~(\ref{composition}) above, the theory ${\sf SLat} \cup {\sf
  Mon}_{\Sigma} \cup \K$ of semilattices with
monotone operators satisfying axioms $\K$  is $\leq$-interpolating
with the notion of $\Theta_\K$-sharing for uninterpreted function symbols.  
\label{thm:interpol-SLO}
\end{thm}
{\em Proof.} The clauses of type (\ref{inclusion})
and~(\ref{composition}) satisfy the conditions 
in the statement of Proposition~\ref{sep} and Theorem~\ref{thm-pint}.  
The result is therefore a consequence of the fact that
the theory extension ${\sf SLat} \subseteq {\sf SLat}  \cup {\sf
  Mon}_{\Sigma} \cup \K$ satisfies conditions ${\bf A1, A2, A3, A4}$ 
and of Proposition~\ref{sep} and 
Theorem~\ref{thm-pint}. \QED

\smallskip
\noindent We illustrate the way Theorem~\ref{lemma-rel-transl}, Proposition~\ref{sep} and Theorem~\ref{thm-pint} 
and the algorithm in Section~\ref{p-int} can be used for computing intermediate terms below: 
\begin{example}
Consider the extension ${\sf SLO} = {\sf SLat} \cup {\sf Mon}_f \cup {\sf Mon}_g
\cup \K$ of ${\sf SLat}$ with two monotone functions $f, g$
satisfying: $\K = \{ y \leq g(x) \rightarrow f(y) \leq g(x) \}$. 
Consider the following conjunctions of atoms: 
$A := d \leq g(a) ~\wedge~ a \leq c ~\wedge~ g(c) \leq a$ and 
$B := b \leq d \wedge b \leq f(b)$. 
It can be checked that $A \wedge B \models b \leq a$. 

To obtain a
separating term we proceed as follows: 
By the definition of ${\sf SLO}$, $A \wedge B \models_{SLO} b \leq a$
iff ${\sf SLat} \wedge {\sf Mon}_f \wedge {\sf Mon}_g \wedge \K \wedge
A \wedge B \wedge \neg( b \leq a) \models \perp$.
By Theorem~\ref{lemma-rel-transl}, 
this is the case iff ${\sf SLat} \wedge ({\sf Mon}_f \wedge {\sf Mon}_g \wedge \K)[\Psi(G)] \wedge
G \models \perp$, where $G = A \wedge B \wedge \neg( b \leq a)$, ${\sf
  est}(G) = \{ g(a), g(c), f(b) \}$ and $\Psi(G) = \{ g(a), g(c), f(b)
\}$.
\begin{itemize}
\item ${\sf
  Mon}_f[\Psi(G)] = \{ b \leq b \rightarrow f(b) \leq f(b) \}$ (redundant). 
\item ${\sf Mon}_g[\Psi(G)] = \{ d_1 \leq d_2 \rightarrow g(d_1) \leq
  g(d_2) \mid d_1, d_2 \in \{ a, c \} \}$. 
\item $\K[\Psi(G)] = \{ b \leq g(a) \rightarrow f(b) \leq g(a),  b
  \leq g(c) \rightarrow f(b) \leq g(c) \}$.
\end{itemize}
{\bf Step 1:} We purify $({\sf Mon}_f \wedge {\sf Mon}_g \wedge \K)[\Psi(G)] \wedge
G$, by  
introducing constants $a_1$ for $g(a)$, $c_1$ for $g(c)$ and $b_1$ for
$f(b)$ and obtain the formula ${\sf Def} \wedge A_0 \wedge B_0 \wedge
{\sf Mon}_0 \wedge {\K}_0$: 

\medskip
\noindent 
$\begin{array}{@{}l@{}l|lll}
\hline
& {\sf Def} & A_0 \wedge B_0 & & {\sf Mon}_0 \wedge {\K}_0 \\
\hline 
D_A\!:\, & a_1 \approx g(a) \wedge c_1 \approx g(c) & A_0:  d \leq a_1 \wedge a \leq
c \wedge c_1 \leq a ~ & {\sf Mon}_A & a \lhd c \rightarrow a_1 \lhd c_1 \\
D_B\!:\, & b_1 \approx f(b) & B_0: b \leq \, d ~ \wedge b \leq b_1  & \K_{\sf
  mix} & b \leq a_1 \rightarrow
b_1 \leq a_1 \\
& & ~~~\lhd \in \{ \leq, \geq \} & & b \leq c_1 \rightarrow
b_1 \leq c_1 \\
\end{array}$

\medskip
\noindent The instances of the congruence axioms ${\sf Con}[D_A \wedge
D_B]$ are redundant in the presence of the corresponding instances of
the monotonicity axioms for $f$ and $g$ and can therefore be ignored.
 
\medskip
\noindent {\bf Step 2.} $\Delta := \top$. 
Find clauses in ${\sf Mon}_0 \wedge {\K}_0$
with premises entailed by $A_0 \wedge B_0 \wedge \Delta$. 

\begin{description} 
\item[$C = a \leq c \rightarrow a_1 \leq c_1$:] $C$ is not mixed.
Since $A_0 \wedge B_0 \models_{\sf SLat} a
\leq c$, $A_0 \wedge B_0 \wedge (a \leq c \rightarrow a_1
\leq c_1)$ is equivalent to $A_0 \wedge B_0 \wedge a_1 \leq c_1$. 
Let $\Delta := \{ a_1 \leq c_1 \}$. 

\item[$C = b \leq a_1 \rightarrow b_1 \leq a_1$:] $C$ is mixed. 
Since $A_0 \wedge B_0 \wedge a_1 \leq c_1 \models b \leq a_1$ 
we find a separating term. For this we use the method
described in the proof of Lemma~\ref{slat-p-interp}. 
We consider the encoding  $N_B \wedge P_b := (P_b \rightarrow P_d)
\wedge (P_b \rightarrow P_{b_1}) \wedge P_b$. Using ordered resolution 
with an ordering in which $P_b, P_{b_1} \succ P_d$ we derive the unit
clauses $P_d$ and $P_{b_1}$. Since $d$ is the only shared constant, $t
= d$ is the separating term. 
Thus,  $A_0 \wedge B_0 \wedge a_1 \leq c_1
\models b \leq d ~\wedge~ d \leq a_1$. We now can separate the
instance $b \leq a_1 \rightarrow b_1 \leq a_1$ of the clause in $\K$
by introducing a new shared constant $d_1$ as a name for $f(d)$ and 
replacing the clause, as
described in the algorithm at the end of Section~\ref{p-int}, with the conjunction of 
\begin{itemize}
\item[(i)] $b \leq d \rightarrow b_1 \leq d_1$ \quad ~~(corresponding to
  $b \leq d \rightarrow f(b) \leq f(b)$) 
\item[(ii)] $d \leq a_1
\rightarrow d_1 \leq a_1$ \quad (corresponding to $d
\leq g(a) \rightarrow f(d) \leq g(a)$)
\end{itemize}
((i) is an instance of a monotonicity
axiom, (ii) is another instance of $\K$), 
and 
 $A_0 \wedge B_0  \wedge a_1 \leq c_1 \wedge (b \leq d \rightarrow b_1
 \leq d_1) \wedge (d \leq a_1
\rightarrow d_1 \leq a_1)$ is equivalent to $A_0 \wedge B_0 \wedge a_1
\leq c_1 \wedge b_1 \leq d_1 \wedge d_1 \leq a_1$. 
Let $\Delta := \Delta \wedge b_1 \leq d_1 \wedge d_1 \leq a_1$.
\end{description}
{\bf Step 3:} The last conjunction entails $b \leq a$. 
To compute a separating term, we again use Lemma~\ref{slat-p-interp}. 
We consider the encoding $N'_B \wedge P_b := (P_b \rightarrow P_d)
\wedge (P_b \rightarrow P_{b_1}) \wedge (P_{b_1} \rightarrow P_{d_1}) \wedge
P_b$ of the $B$-part of the conjunction, $B_0 \wedge b_1 \leq d_1$. 
Using ordered resolution 
with an ordering in which $P_b, P_{b_1} \succ P_d, P_{d_1}$ we derive the unit
clauses $P_d, P_{b_1}$ and $P_{d_1}$. Since $d, d_1$ are the shared constants, $t
= d \swedge d_1$ is the separating term w.r.t.\ ${\sf SLat}$. 
Therefore, $d \swedge f(d)$ is
a separating term w.r.t.\ ${\sf SLO}$. 
(it can in fact be seen that already $d$ is a separating term). \QED
\end{example}
\begin{thm}
If $\K$ contains axioms of type~(\ref{composition}) then the theory of semilattices
with operators is not $\leq$-interpolating when sharing is regarded as
{\em intersection-sharing}. 
\end{thm}
{\em Proof:} Indeed, assume that for every $\K$ containing 
axioms of type~(\ref{composition}), ${\sf SLat}_{\Sigma}(\K)$ is
$\leq$-interpolating w.r.t.\ intersection-sharing. Then it would also
be $\approx$-interpolating w.r.t.\ intersection-sharing. This cannot be
the case, as can be seen from Example~\ref{example-not-int} which 
is presented in what follows. \QED 

\begin{example}
Consider the theory ${\sf SLat}_{\Sigma}(\K)$ of semilattices with monotone operators $f, g$
satisfying the axioms $\K = \{ x \leq g(y) \rightarrow f(x) \leq g(y)
\}$, and let $C$ be a set of constants containing constants $a, b, 
d, e$.
We show that this theory does not have the
$\Sigma_S$-Beth-definability property, where $\Sigma_S = \{ g, e \}$.

Consider the conjunction of literals $A = (a \leq f(e)) \wedge (e \leq
g(b)) \wedge (g(b) \leq a)$. One can prove that $a$ is implicitly
definable w.r.t.\ $\{ g, e \}$ by proving, using the hierarchical
reduction for local theory extensions in Theorem~\ref{lemma-rel-transl},
that: 

\smallskip
\noindent 
{\small $(a {\leq} f(e)) \wedge (e {\leq} g(b)) \wedge (g(b) {\leq} a) \wedge
(a' {\leq} f'(e)) \wedge (e {\leq} g(b')) \wedge (g(b') {\leq} a') 
{\models}_{{\sf Slat}_{\Sigma}(\K {\cup} \K')} a {\approx} a'.$
}

\smallskip
\noindent We show that 
$a$ is not explicitly definable w.r.t.\  $\{ g, e \}$. 
If there exists a term $t$ containing only $g$ and $e$ such that 
$(a {\leq} f(e)) \wedge (e {\leq} g(b)) \wedge (g(b) {\leq} a)
\models_{{\sf Slat}_{\Sigma}(\K)} a {\approx} t$, 
then the interpretations of $a$ and $t$ are equal in every model of
${\sf SLat}_{\Sigma}(\K)$ which is a model of $A$. We show that this is
not the case. 

\smallskip
\noindent 
Let $S = (\{ a_S, e_S, b_S, d_S\}, \swedge, f_S, g_S)$ be the semilattice 
where:
\begin{itemize}
\item $d_S \leq e_S \leq a_S$, $d_S \leq b_S$ and  $a_S \swedge b_S = e_S \swedge b_S = d_S$, 
\item $f_S(a_S) = f_S(e_S) = a_S$, $f_S(b_S) = f_S(d_S) = d_S$, 
\item $g_S(a_S) = g_S(e_S) = g_S(d_S) = d_S$ and $g_S(b_S) = a_S$. 
\end{itemize}
Then $S$ satisfies $A$, $f_S$ and $g_S$ are monotone. 
We prove that $S$ is a model of $\K$: 
Let $x, y \in S$. Assume that $x \leq g_S(y)$. We show that $f_S(x)
\leq g_S(y)$.
\begin{itemize}
\item If $y \in \{ a_S, e_S, d_S \}$ then $g_S(y) = d_S$ so
$x = d_S$, and $f_S(d_S) = d_S \leq g_S(y)$. 
\item If $y = b_S$ then $g_S(b_S) = a_S$, so $x$ can be $a_S, e_S$ or $d_S$, and $f_S(a_S) =
f_S(e_S) = a_S$, $f_S(d_S) = d_S$, so  $f_S(x) \leq g_S(b_S) = a_S$. 
\end{itemize}
A term $t$ containing only $g$ and $e$ can be $e$ or can contain
occurrences of $g$.  If $t = e$ then the interpretation of $t$ in $S$
is $e_S \neq a_S$. If $t$ contains occurrences of $g$ it can be proven that
the interpretation of $t$ in $S$ is $d_S$, i.e.\ is again different from $a_S$. 

\smallskip
\noindent Thus $\T = {\sf SLat}_{\Sigma}(\K)$ does not have the 
Beth definability property w.r.t.\ $\Sigma_S$, hence, by Theorem~\ref{convex-implies-beth}, 
$\T \cup \T' = {\sf SLat}_{f, g}(\K) \cup {\sf SLat}_{f', g}(\K') =
{\sf SLat}_{f,f',g}(\K \cup \K')$, where $\K' = \{ y \leq g(x) \rightarrow
f'(y) \leq g(x) \}$, does not have the $\approx$-interpolation
property w.r.t.\ intersection-sharing, hence it does not have the $\leq$-interpolation property
w.r.t.\ intersection-sharing. 

\smallskip
\noindent {\bf Remark:} By Theorem~\ref{thm:interpol-SLO} and Theorem~\ref{convex-implies-beth},
$\T$ has the $\Theta_\K(\Sigma_S)$-Beth definability property, where $\Theta_\K(\Sigma_S) =
\{ f, g, e \}$. Indeed, then $A \models a \approx f(e)$. 
\QED
\label{example-not-int}
\end{example}
%
\section{Applications to ${\cal EL}$ and ${\cal EL}^+$-Subsumption}
\label{el}

We now explain how these results can be used in the study
of the description logics ${\cal EL}$ and ${\cal EL}^+$. 
In any description logic a set $N_C$ of 
{\em concept names} and a set $N_R$ of {\em roles} is assumed to be
given.
{\em Concept descriptions} can be defined  
with the help of a set of {\em concept constructors}. 
The available constructors determine the expressive power of a description
logic. 
If we only allow intersection and existential restriction
as concept constructors, we obtain the description logic 
${\cal E}{\cal L}$ \cite{Baader2003}, a logic used in 
terminological reasoning in medicine \cite{snomed1,snomed2}. 
The table below shows the constructor names used in 
${\cal E}{\cal L}$ and their semantics. 

\medskip
{\small 
~~~~~~\begin{tabular}{|l|l|l|}
\hline
Constructor name & Syntax & Semantics \\
\hline
\hline
conjunction & $C_1 \sqcap C_2$ & $C_1^{\cal I} \cap C_2^{\cal I}$ \\
\hline
existential restriction & $\exists r.C$ & $\{ x \mid \exists y ((x,y) \in 
r^{\cal I} \mbox{  and } y \in  C^{\cal I}) \}$ \\
\hline
\end{tabular}
}

\medskip
\noindent The semantics is given by interpretations ${\cal I} =
(\Delta, \cdot^{\cal I})$, where $C^{\cal I} \subseteq \Delta$ and
$r^{\cal I} \subseteq \Delta^2$ for every $C \in N_C$, 
$r \in N_R$. The extension of $\cdot^{\cal I}$ to concept descriptions is 
inductively defined using the semantics of the constructors.
In \cite{Baader-2005,Baader-dl-2006}, the 
extension ${\cal E}{\cal L}^+$ of ${\cal E}{\cal L}$ 
with role inclusion axioms is studied. 

\begin{definition}
A TBox (or terminology) is a finite 
set consisting of 
{\em general concept inclusions} (GCI) of the form 
$C \sqsubseteq D$, where $C$ and $D$ are concept descriptions. 
A CBox consists of a TBox  and a set of role inclusions 
of the form $r_1 \circ \dots \circ r_n \sqsubseteq s$, so 
we view CBoxes as 
unions $GCI {\cup} {\cal R}$ of a set $GCI$ of general concept inclusions and 
a set ${\cal R}$ of role inclusions 
of the form $r_1 \circ \dots \circ r_n \sqsubseteq s$, with 
$n {\geq} 1$.\footnote{It can be shown that it is sufficient to consider role inclusions of the form 
$r \sqsubseteq s$ or $r_1 \circ r_2 \sqsubseteq s$, where $r, s, r_1,
r_2$ are role names \cite{Baader-2005}.}
\end{definition}
\begin{definition}
An interpretation ${\cal I}$ is a {\em  model 
of the CBox} ${\cal C} = GCI \cup {\cal R}$ if it is a model of $GCI$,  
i.e., $C^{\cal I} {\subseteq} D^{\cal I}$ for every 
$C {\sqsubseteq} D \in GCI$,  
and 
satisfies all role inclusions in ${\cal C}$, i.e.,  
$r_1^{\cal I} \circ \dots \circ r_n^{\cal I} \subseteq s^{\cal I}$ 
for all $r_1 \circ \dots \circ r_n \subseteq s \in {\cal R}$.
If ${\cal C}$ is a CBox and $C_1, C_2$ are concept descriptions, 
then ${\cal C} \models C_1 \sqsubseteq C_2$  
if and only if $C_1^{\cal I} \subseteq C_2^{\cal I}$ for every model  
${\cal I}$ of ${\cal C}$.
\end{definition}

\subsection{Algebraic semantics for ${\cal EL}, {\cal EL}^+$ and
  $\sqsubseteq$-interpolation}

\noindent In \cite{Sofronie-amai-07} we studied the link between 
TBox subsumption in  $\cal{EL}$ 
and uniform word problems in the corresponding classes 
of semilattices with monotone functions. 
In \cite{sofronie-aiml08}, we showed 
that these results naturally extend to CBoxes and to the description logic
$\cal{EL}^+$. 
When defining the semantics of ${\cal EL}$ or ${\cal EL}^+$ with role
names $N_R$ we use 
a class of $\swedge$-semilattices with monotone operators of the form 
 ${\sf SLat}_{\Sigma}$, where $\Sigma = \{ f_r \mid r \in N_R \}$. 
Every concept description $C$ can be represented as a term ${\overline
  C}$; the encoding is inductively defined:
\begin{itemize}
\item Every concept name $C \in N_C$ is regarded as a constant
  ${\overline C} = C$. 
\item  $\overline{C_1 \sqcap C_2} := {\overline C_1} \swedge {\overline
    C_2}$, and  
\item $\overline{\exists r C} = f_{r}({\overline C})$.
\end{itemize}
If ${\cal R}$ is a set of  role inclusions of the form $r \sqsubseteq s$ and 
$r_1 \circ r_2 \sqsubseteq s$, let $\K$ be the set of all axioms 
of the form: 

\medskip
$ \begin{array}{rll}
\forall x & (f_{r}(x) \leq f_{s}(x))  & \text{ for all
} r \sqsubseteq s \in {\cal R} \\
\forall x & (f_{r_1}( f_{r_2}(x)) \leq f_{s}(x))  & \text{ for all
} r_1 \circ r_2 \sqsubseteq s \in {\cal R} 
\end{array}$
\begin{thm}[\cite{sofronie-aiml08}] 
Assume that the only concept constructors are 
intersection and existential restriction. Then  
for all concept descriptions $D_1, D_2$ and every $\cal{EL}^+$ CBox 
${\cal C} {=} GCI {\cup} {\cal R}$ -- where ${\cal R}$ consists of role
inclusions of the form $r \sqsubseteq s$ and 
$r_1 \circ r_2 \sqsubseteq s$ -- 
with concept names 
$N_C = \{ C_1, \dots, C_n \}$ and set of roles $N_R$:  
$${\cal C} \models D_1 {\sqsubseteq} D_2 \quad \text{ iff }   \quad 
\left( \bigwedge_{C {\sqsubseteq} D \in GCI} 
\overline{C} {\leq} \overline{D} \right) \models_{{\sf SLat}_{\Sigma}(\K)} \overline{D_1}
{\leq} \overline{D_2},$$
where $\Sigma$ is associated with $N_R$ and $\K$ with ${\cal R}$ as
described above.
\label{el-slat}
\end{thm}


\noindent In \cite{ten-Cate-et-al-13,kr-22} the following notion of interpolation, which we
call $\sqsubseteq$-interpolation, is defined: 
A description logic has the $\sqsubseteq$-interpolation property if 
for any CBoxes ${\cal C}_A = GCI_A \cup {\cal R}_A$, ${\cal C}_B =
GCI_B \cup {\cal R}_B$ and any concept descriptions $C, D$ 
such that ${\cal C}_A \cup {\cal C}_B \models C \sqsubseteq D$ there exists
a concept description $T$ containing only concept and role 
symbols ``shared'' by $\{ {\cal C}_A, C \}$ and $\{ {\cal C}_B, D \}$
such that ${\cal C}_A \cup {\cal C}_B \models C \sqsubseteq T$ and 
${\cal C}_A \cup {\cal C}_B \models T \sqsubseteq D$. 
By Theorem  \ref{el-slat}, ${\cal C}_A \cup {\cal C}_B \models C
\sqsubseteq D$ if and only if  
$A \wedge B  \models_{{\sf
  SLat}_{\Sigma}(\K)} \overline{C}
{\leq} \overline{D},$ where $A = \bigwedge_{C_1 {\sqsubseteq} C_2 \in GCI_A} 
\overline{C_1} {\leq} \overline{C_2}$ and $B = \bigwedge_{C_1 {\sqsubseteq} C_2 \in GCI_B} 
\overline{C_1} {\leq} \overline{C_2}$, and 
$\K  = \K_A \cup \K_B$, the union of the axioms associated
with the set inclusions ${\cal R}_A$ resp. ${\cal R}_B$.
By Theorem~\ref{thm:interpol-SLO}, there exists a term 
containing only constants and function symbols 
{\em $\Theta_{\K_A \cup \K_B}$-shared} by 
$A$ and $B$ such that $A \wedge B \models_{{\sf
  SLat}_{\Sigma}(\K_A \cup \K_B)} \overline{C}
{\leq} t \wedge t {\leq} \overline{D}.$ From $t$ we can construct a
concept description $T$ containing only concept names and roles {\em
  shared} by ${\cal C}_A$ and ${\cal C}_B$, and by Theorem~\ref{el-slat},
$C_A \wedge C_B \models C \sqsubseteq T \wedge T \sqsubseteq D$. 
Therefore, the $\sqsubseteq$-interpolation problem studied for  description
logics in \cite{ten-Cate-et-al-13,kr-22} can be expressed in the case
of ${\cal EL}$ and ${\cal EL}^+$ as a
$\leq$-interpolation problem in the class of semilattices with
operators, and the hierarchical method for $\leq$-interpolation 
can be used in this case. We distinguish between intersection-sharing 
and $\Theta_{\cal R}$-sharing, where $\Theta_{\cal R}$ is
the analogon of $\Theta_\K$ where $\K$ is the translation of ${\cal R}$.
\begin{cor}
${\cal EL}$ and ${\cal EL}^+$ 
have the $\sqsubseteq$-interpolation property w.r.t.\ 
  $\Theta_{\cal R}$-sharing. 
\end{cor}
\begin{cor}
${\cal EL}^+$ 
with role inclusions of the form $r_1 \circ r_2 \sqsubseteq s$ 
does not have $\sqsubseteq$-interpolation w.r.t.\
intersection-sharing.
\end{cor}

\subsection{Example: $\sqsubseteq$-Interpolation for ${\cal EL}^+$}
\label{app:p-int}

We now explain our method in detail and illustrate each
step of the method for $\leq$-interpolation described in Section~\ref{p-int}
on an example.

\begin{example}
\label{ex:el3}

\begin{figure}[t]
\fbox{\parbox{0.98\textwidth}{
\vspace{-2mm}
\begin{align*}
~A_1 &:&	~~~~\textsf{Endocardium} ~~&\sqsubseteq~~ \textsf{Tissue}\\
~A_2 &:&	~~~~\textsf{Endocardium} ~~&\sqsubseteq~~ \exists\textsf{part-of.HeartWall}\\
~A_3 &:&	~~~~\textsf{HeartWall} ~~&\sqsubseteq~~ \textsf{BodyWall}\\
~A_4 &:&	~~~~\textsf{HeartWall} ~~&\sqsubseteq~~ \exists\textsf{part-of.LeftVentricle}\\
~A_5 &:&	~~~~\textsf{HeartWall} ~~&\sqsubseteq~~ \exists\textsf{part-of.RightVentricle}\\
~A_6 &:&	~~~~\textsf{LeftVentricle} ~~&\sqsubseteq~~ \textsf{Ventricle}\\
~A_7 &:&	~~~~\textsf{RightVentricle} ~~&\sqsubseteq~~ \textsf{Ventricle}\\
~A_8 &:&	~~~~\textsf{Endocarditis} ~~&\sqsubseteq~~ \textsf{Inflammation}\\
~A_9 &:&	~~~~\textsf{Endocarditis} ~~&\sqsubseteq~~ \exists\textsf{has-location.Endocardium}~~~~\\
~A_{10} &:&	~~~~\textsf{Inflammation} \sqcap \exists\textsf{has-location.Endocardium} ~~&\sqsubseteq~~ \textsf{Endocarditis}\\
~A_{11} &:&	~~~~\textsf{Inflammation} ~~&\sqsubseteq~~ \textsf{Disease}\\
~A_{12} &:&	~~~~\textsf{Inflammation} ~~&\sqsubseteq~~ \exists\textsf{acts-on.Tissue}\\
\\
~B_1 &:&	~~~~\textsf{Ventricle} ~~&\sqsubseteq~~ \exists\textsf{part-of.Heart}\\
~B_2 &:&	~~~~\textsf{HeartDisease} ~~&\sqsubseteq~~ \textsf{Disease}\\
~B_3 &:&	~~~~\textsf{HeartDisease} ~~&\sqsubseteq~~ \exists\textsf{has-location.Heart}\\
~B_4 &:&	~~~~\textsf{Disease} \sqcap \exists\textsf{has-location.Heart} ~~&\sqsubseteq~~ \textsf{HeartDisease}\\
\\
~R_1 &:&	~~~~\textsf{part-of} \circ \textsf{part-of} ~~&\sqsubseteq~~ \textsf{part-of}\\
~R_2 &:&	~~~~\textsf{has-location} \circ \textsf{part-of} ~~&\sqsubseteq~~ \textsf{has-location}
\end{align*}
}}
\caption{Ontology ${\cal O}_{Med}$}
\label{fig:el-ex3}
\end{figure}

Consider the ontology ${\cal O}_{Med}$ in Figure \ref{fig:el-ex3}.
It is based on an example from \cite{Suntisrivaraporn2009}, 
which we modified in some points. 
We changed the CBox in order to ensure that it only contains general 
concept inclusions and that conjunction only appears 
on the left hand side of an axiom. 
Furthermore we left out some axioms and concepts, but also added new concepts 
($\textsf{LeftVentricle}$, $\textsf{RightVentricle}$, $\textsf{Ventricle}$) and 
changed some axioms accordingly. 
 
We divided the CBox into three parts: The $A$-part is
our main TBox, ${\cal T}_A$, which is supposed to be consistent. 
The $B$-part, TBox ${\cal T}_B$, is an extension of the main CBox and may introduce 
some new (and in the worst case even unwanted) consequences. 
The $R$-part contains only role axioms ${\cal R}$. 

We have the
following sets of symbols (we indicate also the abbreviations used in
what follows): 
\begin{align*}
N^A_{C} &=& \{ &\textsf{Endocardium (Em), Tissue (T), HeartWall (HW),} \\
 			&&&\textsf{LeftVentricle (LV), RightVentricle
                          (RV), Ventricle (V)}, \\
& & & \textsf{Disease (D), Inflammation (I), Endocarditis (Es)} \} \\	
N^B_{C} &=& \{ &\textsf{Heart (H), HeartDisease (HD), Disease (D),
  Ventricle (V)} \} \\		
N^{AB}_{C} &=& \{ &\textsf{Disease (D), Ventricle (V)} \}
\end{align*}
Consider the subsumption $\textsf{Endocarditis} \sqsubseteq \textsf{HeartDisease}$. We have $\textsf{Endocarditis} \in N^A_{C}$ and $\textsf{HeartDisease} \in N^B_{C}$ and additionally the following hold: 
\begin{align*} 
{\cal T}_A \cup {\cal T}_B \cup {\cal R} &\models \textsf{Endocarditis}
\sqsubseteq \textsf{HeartDisease} 
\end{align*}
and, in addition:
$${\cal T}_A \cup {\cal R} \not\models \textsf{Endocarditis} \sqsubseteq
\textsf{HeartDisease} \quad 
{\cal T}_B \cup {\cal R} \not\models \textsf{Endocarditis} \sqsubseteq \textsf{HeartDisease}$$
Therefore, we can use the method described in Section~\ref{p-int}
(based on Proposition~\ref{sep}, Theorem~\ref{thm-pint} and Lemma~\ref{slat-p-interp})
to compute an intermediate term containing only shared symbols for the subsumption $\textsf{Endocarditis} \sqsubseteq \textsf{HeartDisease}$, which serves as an explanation for the subsumption. 
\begin{figure}[t]
\fbox{\parbox{0.98\textwidth}{
\begin{minipage}{0.43\textwidth}
\vspace{-3mm}
\begin{align*}
A_1 &:&	~\textsf{Em} ~~&\leq~~ \textsf{T}\\
A_2 &:&	~\textsf{Em} ~~&\leq~~ \textsf{po(HW)}\\
A_3 &:&	~\textsf{HW} ~~&\leq~~ \textsf{BW}\\
A_4 &:&	~\textsf{HW} ~~&\leq~~ \textsf{po(LV)}\\
A_5 &:&	~\textsf{HW} ~~&\leq~~ \textsf{po(RV)}\\
A_6 &:&	~\textsf{LV} ~~&\leq~~ \textsf{V}\\
A_7 &:&	~\textsf{RV} ~~&\leq~~ \textsf{V}\\
A_8 &:&	~\textsf{Es} ~~&\leq~~ \textsf{I}\\
A_9 &:&	~\textsf{Es} ~~&\leq~~ \textsf{hl(Em)}~~~~\\
A_{10} &:&	~\textsf{I} \wedge \textsf{hl(Em)} ~~&\leq~~ \textsf{Es}\\
A_{11} &:&	~\textsf{I} ~~&\leq~~ \textsf{D}\\
A_{12} &:&	~\textsf{I} ~~&\leq~~ \textsf{ao(T)}
\end{align*}
\end{minipage}
\begin{minipage}{0.53\textwidth}
\begin{align*}
B_1 &:&	~\textsf{V} ~~&\leq~~ \textsf{po(H)}\\
B_2 &:&	~\textsf{HD} ~~&\leq~~ \textsf{D}\\
B_3 &:&	~\textsf{HD} ~~&\leq~~ \textsf{hl(H)}\\
B_4 &:&	~\textsf{D} \wedge \textsf{hl(H)} ~~&\leq~~ \textsf{HD}\\
\\
R_1 &:&	~\forall\textsf{X: po(po(X))} ~~&\leq~~ \textsf{po(X)}\\
R_2 &:&	~\forall\textsf{X: hl(po(X))} ~~&\leq~~ \textsf{hl(X)}\\
\\
M_1 &:& ~\forall\textsf{X,Y:~~ X} \leq \textsf{Y} ~~&\rightarrow~~ \textsf{po(X)} \leq \textsf{po(Y)}\\
M_2 &:&	~\forall\textsf{X,Y:~~ X} \leq \textsf{Y} ~~&\rightarrow~~ \textsf{hl(X)} \leq \textsf{hl(Y)}\\
M_3 &:&	~\forall\textsf{X,Y:~~ X} \leq \textsf{Y} ~~&\rightarrow~~ \textsf{ao(X)} \leq \textsf{ao(Y)}
\end{align*}
\end{minipage}
}}
\caption{${\cal O}_{Med}$ after translation to \textsf{SLat} with monotone operators}
\label{step1}
\end{figure}

\medskip
\noindent \textbf{Step 1:} We translate the original ontology to the
theory of semilattices with operators. 
We now state the monotonicity axioms for each role explicitly.
Figure \ref{step1} shows the ontology after the translation to the
theory of semilattices with operators.
Note that from here on we use the abbreviations for concept names indicated in the sets $N^A_{C}, N^B_{C}$ and $N^{AB}_{C}$ above
and also abbreviations for role names, i.e.\ $\textsf{po}$ for $\textsf{part-of}$, $\textsf{hl}$ for $\textsf{has-location}$ and $\textsf{ao}$ for $\textsf{acts-on}$.

\medskip
\noindent \textbf{Step 2:} Using unsat core computation we get the following minimal axiom set:

~~~~~~~$min_A=\{A_2, A_4, A_6, A_8, A_9, A_{11}, B_1, B_4, R_2 \}$ 

\noindent 
This means that for the following instantiation step we only have to consider the role axiom $R_2$ and none of the monotonicity axioms is needed.

\medskip
\noindent \textbf{Step 3:} Let ${\cal T}_0 = \textsf{SLat}$ and 
$T_1 = \textsf{SLat} \cup R_2$ be the extension of $T_0$ with axiom
$R_2$. 
We know that it is a local theory extension, so we can use
hierarchical reasoning. 
We first flatten the role axiom $R_2$ in the following way:

~~~~~~~$R_2^\textsf{flat}: ~~\forall \textsf{X, Y: ~~X} \leq \textsf{po(Y)} ~~\rightarrow~~ \textsf{hl(X)} \leq \textsf{hl(Y)}$

\noindent We have the following set of ground terms:

~~~~~~~$G = \textsf{est}(\T_A \cup \T_B \cup {\cal R}) = \{~\textsf{po(HW)},~ \textsf{po(LV)},~ \textsf{po(H)},~ \textsf{hl(Em)},~ \textsf{hl(H)}~ \}$

\noindent We use the closure operator $\Psi$ described in
Lemma~\ref{lemma-psi} to extend our set of ground terms: For every term $\textsf{po(X)}$ 
in $G$ we have to add the term $\textsf{hl(X)}$ and vice versa. This leads to the following extended set $G'$ of ground terms:

~~~~~~~$G' = \{~\textsf{po(Em)},~ \textsf{po(HW)},~ \textsf{po(LV)},~ \textsf{po(H)},~ \textsf{hl(Em)},~ \textsf{hl(HW)},~ \textsf{hl(LV)},~ \textsf{hl(H)}~ \}$

\noindent From $G'$ we get the following instances of the axiom $R_2^\textsf{flat}$:

{\small \begin{minipage}{0.6\textwidth}
\vspace{-3mm}
\begin{align*}
~~~I_1:& & \textsf{Em} &\leq \textsf{po(HW)}& &\rightarrow& \textsf{hl(Em)} &\leq \textsf{hl(HW)}\\
~~~I_2:& & \textsf{Em} &\leq \textsf{po(LV)}& &\rightarrow& \textsf{hl(Em)} &\leq \textsf{hl(LV)}\\
~~~I_3:& & \textsf{Em} &\leq \textsf{po(H)}& &\rightarrow& \textsf{hl(Em)} &\leq \textsf{hl(H)}\\
~~~I_4:& & \textsf{HW} &\leq \textsf{po(Em)}& &\rightarrow& \textsf{hl(HW)} &\leq \textsf{hl(Em)}\\
~~~I_5:& & \textsf{HW} &\leq \textsf{po(LV)}& &\rightarrow& \textsf{hl(HW)} &\leq \textsf{hl(LV)}\\
~~~I_6:& & \textsf{HW} &\leq \textsf{po(H)}& &\rightarrow& \textsf{hl(HW)} &\leq \textsf{hl(H)}\\
~~~I_7:& & \textsf{LV} &\leq \textsf{po(Em)}& &\rightarrow& \textsf{hl(LV)} &\leq \textsf{hl(Em)}\\
~~~I_8:& & \textsf{LV} &\leq \textsf{po(HW)}& &\rightarrow& \textsf{hl(LV)} &\leq \textsf{hl(HW)}\\
~~~I_9:& & \textsf{LV} &\leq \textsf{po(H)}& &\rightarrow& \textsf{hl(LV)} &\leq \textsf{hl(H)}\\
~~~I_{10}:& & \textsf{H} &\leq \textsf{po(Em)}& &\rightarrow& \textsf{hl(H)} &\leq \textsf{hl(Em)}\\
~~~I_{11}:& & \textsf{H} &\leq \textsf{po(HW)}& &\rightarrow& \textsf{hl(H)} &\leq \textsf{hl(HW)}\\
~~~I_{12}:& & \textsf{H} &\leq \textsf{po(LV)}& &\rightarrow& \textsf{hl(H)} &\leq \textsf{hl(LV)}
\end{align*}
\end{minipage}
} 

\smallskip
\noindent 
We purify all formulae by introducing new constants for the terms starting with a function symbol, i.e.\ role names. We save the definitions in the following set:
\begin{align*}
\textsf{Def} = \{ &\textsf{po}_\textsf{HW}=\textsf{po(HW)}, \textsf{po}_\textsf{LV}=\textsf{po(LV)}, \textsf{po}_\textsf{H}=\textsf{po(H)}, \textsf{hl}_\textsf{EM}=\textsf{hl(EM)}, \\
&\textsf{hl}_\textsf{HW}=\textsf{hl(HW)}, \textsf{hl}_\textsf{LV}=\textsf{hl(LV)}, \textsf{hl}_\textsf{HC}=\textsf{hl(HC)}, \textsf{hl}_\textsf{H}=\textsf{hl(H)}\}
\end{align*}
We then have the set $A_0 \wedge B_0 \wedge I_0$, where $A_0$, $B_0$ and $I_0$ are the purified versions of $A=\{A_2,A_4,A_6,A_8,A_9,A_{11}\}$, $B=\{B_1,B_4\}$ and $I=\{ I_1,..., I_{10}\}$, respectively.

\medskip
\noindent \textbf{Step 4:} To reduce the number of instances we compute an unsatisfiable core and obtain the following
set of axioms: 

~~~~~~~${min'_A}=\{ A_2, A_4, A_6, A_8, A_9, A_{11}, B_1, B_4, I_1, I_5, I_9 \}$ 

\noindent So we have ${\cal H} = \{I_1,I_5,I_9\}$. Out of these instances the
first two are pure $A$ (meaning the premise contains only symbols in
$N^A_C$), but $I_9$ is a mixed instance, since 
$\textsf{LV} \in N^A_C \backslash N^B_C$ and $\textsf{H} \in N^B_C
\backslash N^A_C$, so ${\mathcal H}_\textsf{mix} = \{I_9\}$.

\medskip
\noindent \textbf{Step 5:} To separate the mixed instance $\textsf{LV} \leq \textsf{po}_\textsf{H} \rightarrow \textsf{hl}_\textsf{LV} \leq \textsf{hl}_\textsf{H}$ 
we use the construction in Proposition~\ref{sep} to compute an intermediate term $t$ in the common signature such that $\textsf{LV} \leq t$ and $t \leq \textsf{po}_\textsf{H}$. We obtain $t = \textsf{V}$. 
We get ${\mathcal H}_\textsf{sep} = \{I_9^A,I_9^B\}$ where:

\begin{minipage}{0.5\textwidth}
\vspace{-3mm}
\begin{align*}
~~~~~~~~~~~~I_9^A:& & \textsf{LV} &\leq \textsf{V}& &\rightarrow& \textsf{hl}_\textsf{LV} &\leq \textsf{hl}_\textsf{V}\\
~~~~~~~~~~~~I_9^B:& & \textsf{V} &\leq \textsf{po}_\textsf{H}& &\rightarrow& \textsf{hl}_\textsf{V} &\leq \textsf{hl}_\textsf{H}
\end{align*}
\end{minipage}

\noindent Note that $I_9^A$ is an instance of the monotonicity axiom for the
$\textsf{has-location}$ role and $I_9^B$ is an instance of axiom
$R_2^\textsf{flat}$.

\medskip
\noindent \textbf{Step 6:} Since for all the instances that are necessary to derive the consequence it must be true that $\T_A \cup \T_B$ entails its premise, it is sufficient to consider only the corresponding conclusions. Note that w.r.t.\ $\textsf{SLat}$ the formula $A_0 \wedge I_1 \wedge I_5 \wedge I_9^A$ is equivalent to the following formula:
{\small \begin{align*}
\overline{A}_0 =~ &\textsf{Em} \leq \textsf{po}_\textsf{HW}
\wedge \textsf{HW} \leq \textsf{po}_\textsf{LV}
\wedge \textsf{LV} \leq \textsf{V}
\wedge \textsf{Es} \leq \textsf{I}
\wedge \textsf{Es} \leq \textsf{hl}_\textsf{Em}
\wedge \textsf{I} \leq \textsf{D}\\
\wedge~ &\textsf{hl}_\textsf{EM} \leq \textsf{hl}_\textsf{HW} 
\wedge \textsf{hl}_\textsf{HW} \leq \textsf{hl}_\textsf{LV} 
\wedge \textsf{hl}_\textsf{LV} \leq \textsf{hl}_\textsf{V}
\end{align*}}
This formula can be seen as a set of Horn clauses ${\overline{A}^h_0}$:
{\small \begin{align*}
\overline{A}^h_0 =~ \{&~
(\neg \textsf{Em} \lor \textsf{po}_\textsf{HW}),~ 
(\neg \textsf{HW} \lor \textsf{po}_\textsf{LV}),~ 
(\neg \textsf{LV} \lor \textsf{V}),~ 
(\neg \textsf{Es} \lor \textsf{I}),~ 
(\neg \textsf{Es} \lor \textsf{hl}_\textsf{Em}),~ 
(\neg \textsf{I} \lor \textsf{D}),~ \\
&~(\neg \textsf{hl}_\textsf{EM} \lor \textsf{hl}_\textsf{HW}),~ 
(\neg \textsf{hl}_\textsf{HW} \lor \textsf{hl}_\textsf{LV}),~ 
(\neg \textsf{hl}_\textsf{LV} \lor \textsf{hl}_\textsf{V}) ~\}
\end{align*}}
To obtain an explanation for ${\cal T}_A \cup
{\cal T}_B \cup {\cal R} \models \textsf{Endocarditis}
\sqsubseteq 
\textsf{HeartDisease}$ 
we saturate the set $\overline{A}^h_0 \cup \{\textsf{Es}\}$ under ordered resolution as described in the proof of
Theorem~\ref{slat-p-interp}, where symbols occurring in $A$ but not in $B$ are larger than common
symbols:
\begin{itemize}
\item Resolution of $\textsf{Es}$ and $\neg \textsf{Es} \lor \textsf{I}$ yields $\textsf{I}$.
\item Resolution of $\textsf{I}$ and $\neg \textsf{I} \lor \textsf{D}$ yields $\textsf{D}$.
\item Resolution of $\textsf{Es}$ and $\neg \textsf{Es} \lor \textsf{hl}_\textsf{Em}$ yields $\textsf{hl}_\textsf{Em}$.
\item Resolution of $\textsf{hl}_\textsf{Em}$ and $\neg \textsf{hl}_\textsf{EM} \lor \textsf{hl}_\textsf{HW}$ yields $\textsf{hl}_\textsf{HW}$.
\item Resolution of $\textsf{hl}_\textsf{HW}$ and $\neg \textsf{hl}_\textsf{HW} \lor \textsf{hl}_\textsf{LV}$ yields $\textsf{hl}_\textsf{LV}$.
\item Resolution of $\textsf{hl}_\textsf{LV}$ and $\neg \textsf{hl}_\textsf{LV} \lor \textsf{hl}_\textsf{V}$ yields $\textsf{hl}_\textsf{V}$.
\end{itemize}
We obtained two resolvents containing only common symbols: $\textsf{D}$ and $\textsf{hl}_\textsf{V}$. 
Taking the conjunction of these terms and translating the formula back to
description logic yields the following formula: 

~~~~~~~~~~~~~~~~~~~~~~~~~$J = \textsf{Disease} ~\sqcap~ \exists\textsf{has-location.Ventricle}$. 

\noindent Indeed, the following properties hold:
\begin{align*}
{\cal T}_A \cup {\cal R} &\models \textsf{Endocarditis} \sqsubseteq J  \\
{\cal T}_A \cup {\cal T}_B \cup {\cal R} &\models ~~~~~~~~~~~~~~~~~~J \sqsubseteq \textsf{HeartDisease} 
\end{align*}

So $J$ is the intermediate term we were looking for. 
\end{example}

\subsection{Prototype implementation}
\smallskip 
\noindent 
The ideas were implemented in a prototype implementation\footnote{The
  implementation and some tests can be found here:\\
  \url{https://userpages.uni-koblenz.de/~sofronie/p-interpolation-and-el/}} 
for the theory of semilattices with operators satisfying axioms of 
type~(\ref{general-form}) considered in this paper. 
The program is written in Python and uses Z3 \cite{z3} and SPASS
\cite{spass} as external provers. 
The program implements {\sf Steps 1-3} in the 
algorithm presented at the end of Section~\ref{p-int} with the following optimization:
In {\sf Step 1} after instantiation and purification, in order to reduce
the size of the set of instances of axioms to be considered, an
unsatisfiable core is computed with Z3. The program separates the mixed instances by
computing intermediate terms for their premises using Theorem~\ref{slat-p-interp}
and Proposition~\ref{sep}; for applying ordered resolution the prover SPASS is used.
In Step~3, the intermediate term $T$ for $C \leq D$ is computed 
using the method described in Theorem~\ref{slat-p-interp}, again using
SPASS. 

For the use for interpolation in $\cal{EL}$ and ${\cal EL}^+$, the
CBoxes ${\cal C}_A$ and ${\cal C}_B$ and the subsumption 
$C \sqsubseteq D$ are given as an input. A minimal subset of ${\cal C}_A \cup {\cal
  C}_B$ is computed from which $C \sqsubseteq D$ can be derived. 
(The user can choose between a precise translation to
SPASS or a propositional translation to Z3 which is not always
precise, but turned out to be a good approximation. Standard implementations available for
computing justifications of entailments from description logic
ontologies could be used as well.) The problem is then translated  into a problem for
$\leq$-interpolation in semilattices with operators. After computing the
interpolating term, the result is expressed in the syntax of description logics. 

\section{Conclusions and future work}
\label{conclusions}

\noindent In this paper we gave a hierarchical method for $P$-interpolation in
certain classes of local theory extensions $\T_0 \subseteq \T_0 \cup
\K$. We used these results for 
proving $\leq$-interpolation in classes of semilattices with monotone
operators satisfying additional clauses $\K$ 
with a suitable notion of $\Theta_\K$-sharing we defined.  
 We defined a form of Beth definability w.r.t. a subsignature
 $\Sigma_S$ and  used it to show that the class of semilattices with
operators under consideration does not have the $\leq$-interpolation property if
only the common function symbols and constants are considered to be
``shared''. 
We discussed how these results can be used for the study of interpolation 
in ${\cal EL}$ and ${\cal EL}^+$. 

\smallskip
\noindent In future work we will explore other application areas of
these results, both to classes of non-classical logics and to theories
relevant in the verification. We will extend the implementation with
possibilities of choosing the base theory and the methods for
$P$-interpolation in the base theory. We will further investigate the
links with Beth definability and possibilities of using Beth
definability for computing explicit definitions for implicitly
definable terms -- and analyze the applicability of such 
results in description logics but also in verification. 

\medskip
\noindent {\bf Acknowledgments.} We thank the reviewers for their
helpful comments. 
The research reported here was funded  by  the  Deutsche
Forschungsgemeinschaft  (DFG,  German  Research Foundation) 
-- Projektnummer 465447331. 

\bibliographystyle{abbrv}
\bibliography{extended-version-cade-2023}

\end{document}